\documentclass[12pt]{amsart}
\usepackage{amsfonts,amssymb,amsthm}
\allowdisplaybreaks
\usepackage[mathscr]{eucal}
\usepackage{hyperref}
\usepackage[matrix,arrow]{xy}
\usepackage{times}

\advance\textheight\topskip
\newcommand{\spaceN}{\mathscr{N}}
\newcommand{\wch}{\mathscr{C}}
\renewcommand{\tilde}{\widetilde}
\renewcommand{\hat}{\widehat}
\newtheorem{prop}{Proposition}[section]
\newtheorem{lemma}[prop]{Lemma}

\renewcommand{\simeq}{\cong}

\newcommand{\T}{\mathrm{T}}
\newcommand{\End}{\mathop{\mathrm{End}}}

\newcommand{\bref}[1]{\textbf{\ref{#1}}}

\newcommand{\dmn}{d}  

\newcommand{\Ker}{\mathop{\mathrm{Ker}}}
\newcommand{\im}{\mathop{\mathrm{Im}}}

\newcommand{\FQ}{C}
\newcommand{\FR}{D}

\newcommand{\fq}{c}
\newcommand{\fr}{d}
\newcommand{\incmap}{\mathfrak{f}}

\newcommand{\p}[1]{|#1|}
\newcommand{\gh}[1]{\mathrm{gh}(#1)}
\newcommand{\momP}{\mathscr{P}}

\newcommand{\Salgebra}{\mathsf{S}}    
\newcommand{\wwedge}{\mbox{\small$\bigwedge$}}

\newcommand{\assalgebra}{\mathscr}    

\newcommand{\assalgA}{\assalgebra{A}}

\newcommand{\algg}{\Liealg{g}}
\newcommand{\algp}{\Liealg{p}}

\newcommand{\Liealg}{\mathfrak}       

\newcommand{\yS}{S}

\newcommand{\map}{\,\mathrm{:}\,}

\newcommand{\dd}{\partial}

\newcommand{\tensor}{\otimes}

\renewcommand{\geq}{\,{\geqslant}\,}

\newcommand{\inner}[2]{\langle #1{,}\,#2\rangle}
\newcommand{\binner}[2]{%
  {\langle}\kern-4.15pt{\langle}#1{,}\,#2{\rangle}\kern-4.15pt{\rangle}}
\newcommand{\commut}[2]{[#1{,}\,#2]}
\newcommand{\pb}[2]{\left\{{}#1{},{}#2{}\right\}}
\newcommand{\ab}[2]{\big(#1,#2\big)}

\newcommand{\half}{\mathchoice{%
    \ffrac{1}{2}}{\frac{1}{2}}{\frac{1}{2}}{\frac{1}{2}}}

\newcommand{\ffrac}[2]{\raisebox{.5pt}%
  {\footnotesize$\displaystyle\frac{#1}{#2}$}\kern1pt}

\newcommand{\brst}{\mathsf{\Omega}}

\newcommand{\st}[2]{\overset{#1}{#2}}
\newcommand{\dr}[1]{\mathchoice{\ffrac{\dd}{\dd #1}}{\frac{\dd}{\dd
      #1}}{\ffrac{\dd}{\dd #1}}{\ffrac{\dd}{\dd #1}}}

\newcommand{\dl}[1]{\mathchoice{\ffrac{\dd}{\dd #1}}{\frac{\dd}{\dd
      #1}}{\ffrac{\dd}{\dd #1}}{\ffrac{\dd}{\dd #1}}}

\newcommand{\ddl}[2]{\ffrac{\dd^L #2}{\dd #1}}

\newcommand{\vddr}[2]{\ffrac{\delta^R #1}{\delta #2}}
\newcommand{\vddl}[2]{{\ffrac{\delta #1}{\delta #2}}}

\newcommand{\vac}{|0\rangle}

\newcommand{\fC}{\mathbb{C}}

\newcommand{\oR}{\mathbb{R}}
\newcommand{\fR}{\mathbb{R}}

\newcommand{\bundle}{\boldsymbol}

\newcommand{\derham}{\boldsymbol{d}}

\newcommand{\manifold}[1]{\mathscr{#1}}
\newcommand{\manX}{\manifold{X}}

\newcommand{\manM}{\manifold{M}}

\def\cC{\mathcal{C}}
\def\cD{\mathcal{D}}
\def\cE{\mathcal{E}}
\def\cF{\mathcal{F}}
\def\cG{\mathcal{G}}
\def\cH{\mathcal{H}}

\def\cL{\mathcal{L}}

\def\cP{\mathcal{P}}

\def\cR{\mathcal{R}}
\def\cS{\mathcal{S}}
\def\cT{\mathcal{T}}
\def\cU{\mathcal{U}}
\def\cV{\mathcal{V}}

\def\cX{\mathcal{X}}

\numberwithin{equation}{section} \makeatletter
\@addtoreset{equation}{section}

\def\@secnumfont{\bfseries}
\def\subsubsection{\@startsection{subsubsection}{3}%
  \z@{.5\linespacing\@plus.7\linespacing}{-.5em}%
  {\normalfont\bfseries}}
\def\paragraph{\@startsection{paragraph}{4}%
  \z@\z@{-\fontdimen2\font}%
  \normalfont\bfseries}
\def\subparagraph{\@startsection{subparagraph}{5}%
  \z@\z@{-\fontdimen2\font}%
  \normalfont\bfseries}

\makeatother

\begin{document}
\pagestyle{myheadings}
\markboth{\textsc{\small Barnich, Grigoriev, Semikhatov, and Tipunin}}{%
  \textsc{\small Parent field theory and unfolding}}
\addtolength{\headsep}{4pt}

\begin{flushright}\small
  ULB-TH/04-17\\[-2pt]
  \texttt{hep-th/0406192}\\[-3pt]
  \texttt{\footnotesize final version}
\end{flushright}

\begin{centering}
  
  \vspace{1cm}  

  \textbf{\Large{Parent field theory and unfolding\\[6pt]
      in BRST first-quantized terms}}
  
  \vspace{1.5cm}
  
  {\large G.~Barnich,$^{*,a}$ \ M.~Grigoriev,$^{\dag,a,b}$ \ 
    A.~Semikhatov,$^b$ \ and I.~Tipunin$^b$}
  
  \vspace{1cm}

  \begin{minipage}{.8\textwidth}\small
    \mbox{}\kern-4pt$^a$Physique Th\'eorique et Math\'ematique and
    International Solvay Institutes, Universit\'e Libre de Bruxelles,
    Campus Plaine C.P. 231, B-1050 Bruxelles, Belgium
    
    \vspace{.5cm}
    
    \mbox{}\kern-4pt$^b$Tamm Theory Department, Lebedev Physics
    Institute, Leninsky prospect 53, 119991 Moscow, Russia
  \end{minipage}

\end{centering}

\vspace{1cm}

\begin{center}
  \begin{minipage}{.9\textwidth}
    \textsc{Abstract}.  For free-field theories associated with BRST
    first-quantized gauge systems, we identify generalized auxiliary
    fields and pure gauge variables already at the first-quantized
    level as the fields associated with algebraically contractible
    pairs for the BRST operator.  Locality of the field theory is
    taken into account by separating the space--time degrees of
    freedom from the internal ones. A standard extension of the
    first-quantized system, originally developed to study quantization
    on curved manifolds, is used here for the construction of a
    first-order parent field theory that has a remarkable property: by
    elimination of generalized auxiliary fields, it can be reduced
    both to the field theory corresponding to the original system and
    to its unfolded formulation. As an application, we consider the
    free higher-spin gauge theories of Fronsdal.
  \end{minipage}
\end{center}

\vfill

\noindent
\mbox{}
\raisebox{-3\baselineskip}{%
  \parbox{\textwidth}{\mbox{}\hrulefill\\[-4pt]}}
{\scriptsize$^*$ Senior Research Associate of the National
  Fund for Scientific Research (Belgium).\\[-2pt]
  $^\dag$ Postdoctoral Visitor of the National Fund for Scientific
  Research (Belgium).}

\thispagestyle{empty} \newpage

\begin{small}
  \tableofcontents
\end{small}

\section{Introduction}
In the context of higher-spin theories, auxiliary fields have been
prominent ever since they have been used by Fierz and Pauli
\cite{Fierz:1939ix} for the construction of Lagrangians for spin-$2$
and $3/2$ fields.  This analysis has been completed in the case of
arbitrary spin by Singh and Hagen \cite{Singh:1974qz,Singh:1974rc}.
The massless limit of these generalized Lagrangians and the emergence
of gauge invariance has then been studied by
Fronsdal~\cite{Fronsdal:1978rb} and by Fang and
Fronsdal~\cite{Fang:1978wz}.

Inspired by string field theory, the Fronsdal Lagrangians, with
additional auxiliary fields, have been reproduced in the mid-eighties
\cite{Ouvry:1986dv,Bengtsson:1986ys,Henneaux:1987cp} as the
Lagrangians associated with a first-quantized BRST system consisting
of a truncation of the tensionless bosonic string.  This reformulation
turns out to be extremely convenient for constructing consistent
interactions, extending to curved backgrounds or studying more general
representations of the Lorentz group (see, e.g.,
\cite{Bengtsson:1988jt,Cappiello:1989cd,Bengtsson:1990un,
  Labastida:1989kw,
  Pashnev:1998ti,Bonelli:2003kh,Bonelli:2003zu,Sagnotti:2003qa,
  Bekaert:2003uc}).

A different framework for higher-spin gauge theories known as the
\textit{unfolded formalism} was developed by
  M.\;Vasiliev~\cite{Vasiliev:1988xc,Vasiliev:1988sa,Vasiliev:1994gr}
  (see also~\cite{Vasiliev:1999ba,Vasiliev:2001zy,Vasiliev:2004qz} for
  a review and further developments).  So far, only in this
approach results on consistent interactions have been obtained to all
orders.  The unfolded formalism is explicitly space--time covariant
and makes both the gauge and global symmetries of the theory
transparent.  This is because the free equations of motion have the
form of a covariant constancy condition, which is achieved at the
price of introducing an infinite tower of auxiliary fields.  In this
formulation, however, the equations do not seem to be naturally
Lagrangian without the elimination or introduction of further
auxiliary fields~\cite{Shaynkman:2000ts}.

The aim of this paper is to explicitly relate the BRST-based approach
and Vasiliev's method of unfolding in the case of free field theories.
For this, we provide a detailed understanding, at the first-quantized
level, of auxiliary fields, pure gauge degrees of freedom, and the
issue of Lagrangian versus non-Lagrangian equations of motion.  We
construct an extended first-order BRST system whose associated field
theory serves as a \textit{parent field theory}: by elimination of
auxiliary fields and pure gauge degrees of freedom, it can be reduced
either to the starting point theory or to its unfolded formulation.

More precisely, we first recall in Sec.~\bref{sec:review} how to
associate a gauge field theory with a BRST first-quantized system
along the lines developed in the context of string field theory.
Technically, the BRST operator of the first-quantized system
determines both the Batalin--Vilkovisky master action and the BRST
differential of the gauge field theory.  Special care is devoted to
locality by treating the space--time degrees of freedom $x^\mu,p_\nu$
separately from internal degrees of freedom.  We also recall that
working on the level of the master action including ghosts and
antifields allows giving a unified description of standard auxiliary
fields and pure gauge degrees of freedom in terms of so-called
\textit{generalized auxiliary fields}.

Section~\bref{sec:NL} begins with the rather obvious remark that for a
BRST first-quantized system, equations of motions, gauge symmetries,
and the BRST differential of an associated field theory can be defined
independently of the question of the existence of a Lagrangian.  More
importantly, for a possibly nonlinear field theory defined by some
differential, we extend the definition of generalized auxiliary fields
to this non-Lagrangian context.

For linear field theories associated with BRST systems, we then
characterize generalized auxiliary fields at the first-quantized level
as the fields associated with \textit{contractible pairs} that can be
eliminated algebraically in the computation of the BRST cohomology.
In the same spirit, existence of a Lagrangian is understood in
first-quantized terms as the existence of an inner product that makes
the BRST operator formally self-adjoint.  These general ideas are then
illustrated by recalling explicitly how the Fronsdal Lagrangians can
be obtained from a first-quantized BRST system.

In Sec.~\bref{sec:fedosov}, we show how to make a given BRST system
first-order in space--time derivatives via a procedure that is
familiar from the problem of quantizing curved manifolds.  The
procedure implies the introduction of new internal degrees of freedom
and an extended BRST operator such that the extended system is
equivalent to the given BRST system.  The extended BRST operator
contains a space--time part and an internal part.  Space--time
derivatives enter only the space--time part, whereby the constraints
describing the embedding of the original phase space into the extended
one are taken into account.  The internal part of the extended BRST
operator coincides with the original BRST operator with space--time
derivatives replaced by derivatives acting in the internal space
enlarged by additional $y$ variables.

We then show that the extended system can be reduced to the original
system by elimination of contractible pairs for the space--time part
of the BRST differential.  Another reduction consists in eliminating
contractible pairs for the internal part of the BRST operator.  The
field theory of this latter reduction is related to the unfolded form
of the equations of motion for the starting point field theory.
Therefore, the field theory of the extended system can be regarded as
a parent theory from which both the original and the unfolded
formulations can be obtained via elimination of generalized auxiliary
fields.  We illustrate the formalism in the case of the Klein--Gordon
equation and the free gauge theories for fields of integer higher
spins.

Finally, we briefly comment on the validity of our analysis in the
case of nonflat backgrounds and discuss symmetries and interactions.
The appendix is devoted to the cohomology of the internal part of the
extended BRST operator for the Fronsdal system.

\section{Local field theory associated with a BRST first-quantized
  system}\label{sec:review}
In this section, we recall some elements of the standard operator BRST
formalism, which we use to describe first-quantized systems of the
type ``particle with internal degrees of freedom.''  Examples of such
systems include the spinless relativistic particle, higher-spin
particle models \cite{Ouvry:1986dv,Bengtsson:1986ys,Henneaux:1987cp},
and strings.  In Sec.~\bref{sec:LFT}, we then consider a gauge field
theory whose fields are associated with the wave functions of the
first-quantized system.  A Lagrangian for this field theory is
constructed using the BRST operator whenever an inner product exists
on the space of first-quantized wave functions.  In
Sec.~\bref{sec:auxiliary}, we recall the concept of generalized
auxiliary fields whose introduction/elimination provides one with a
natural notion of equivalence for such gauge field theories.

\subsection{BRST first-quantized system}\label{subsec:1st-quant}
We consider a constrained Hamiltonian system with phase space
$T^*\manX\times B$, where $T^*\manX$ is the cotangent bundle to a
manifold $\manX$ and $B$ is a symplectic supermanifold.  The
constraints are given by
\begin{equation}\label{constr-ori}
  f_\alpha=0,\qquad \alpha=1,\ldots,s,
\end{equation}
where $f_\alpha$ are functions on $T^*\manX\times B$.  This system can
be considered as a particle on $\manX$ with internal degrees of
freedom described by $B$.

In the BRST approach to constrained
dynamics~\cite{Fradkin:1975cq,Batalin:1977pb,Fradkin:1978xi} (see also
\cite{Henneaux:1985kr}), a ghost $c^\alpha$ and the corresponding
momentum $\momP_\alpha$ must be introduced for each
constraint~$f_\alpha$.  The Grassmann parities of both $c^\alpha$ and
$\momP_\alpha$ are opposite to those of $f_\alpha$.  Let $\Lambda$ be
the linear supermanifold with coordinates $c^\alpha$.  The extended
phase space of the BRST system is then $T^*\manX\times B\times
T^*\Lambda$, with the obvious symplectic structure.  The ghost-number
grading is introduced on functions on $T^*\manX\times B\times
T^*\Lambda$ such that $\gh{c^\alpha}=1$, $\gh{\momP_\alpha}=-1$.

Given the constraints, one constructs the classical BRST charge
$\brst^{cl}$ on the extended phase space, which is a Grassmann-odd and
ghost-number-one function satisfying
\begin{equation}
  \pb{\brst^{cl}}{\brst^{cl}}=0, \qquad
  \brst^{cl}=c^\alpha f_\alpha~+~\text{more}.
\end{equation}
The differential $\pb{\brst^{cl}}{{}\cdot{}}$ provides a resolution of
the algebra of functions on the reduced phase space.  Inequivalent
physical observables then correspond to ghost-number-zero elements of
the cohomology space of $\pb{\brst^{cl}}{{}\cdot{}}$ evaluated in the
algebra of functions on the extended phase space.  The Poisson
structure on the extended phase space naturally induces a Poisson
bracket in cohomology.  We consider two classical BRST systems
equivalent if their ghost-number-zero cohomology spaces are isomorphic
as Poisson algebras.

We next consider the quantization of this system assuming that $\manX$
is $\fR^{\dmn}$.  The states are taken to be sections of the bundle
\begin{equation}
  \bundle{\cH}=\manX\times\cH\to\manX,
\end{equation}
where $\cH$ is the vector (super)space of states arising in the
quantization of $B\times T^*\Lambda$, which we assume to be equipped
with a sesquilinear inner product
$\inner{{}\cdot{}}{{}\cdot{}}_{\cH}$.  The quantum operator algebra
inherits the Grassmann parity and ghost number from the classical
level, but in the representation space $\cH$, both gradings are
defined modulo suspension by a constant.\footnote{For a system without
  physical fermions, the two gradings are compatible in the sense that
  parity corresponds to ghost number modulo~$2$.}  The following
properties of the sesquilinear inner product are assumed:
\begin{equation}
  \overline{\inner{\phi}{\psi}_\cH}=(-1)^{\p{\phi}\p{\psi}}
  \inner{\psi}{\phi}_\cH,
  \qquad
  \inner{\phi\alpha}{\psi\beta}_\cH=\inner{\phi}{\psi}_\cH
  \bar\alpha\beta,\quad \alpha,\beta\subset \fC,
\end{equation}
where $\p{{}\cdot{}}$ denotes Grassmann parity.  We also assume that
an additional grading in $\cH$ can be introduced such that each graded
component is finite-dimensional.

In terms of a real basis $(e_A)$ in $\cH$, the sections of
$\bundle{\cH}$ are written as
\begin{equation}\label{eq:sec1}
  \bundle\phi(x)= e_A\bundle\phi^A(x),
\end{equation}
where $(x^{\mu})$, ${\mu}=1,\dots,\dmn$, are coordinates on $\manX$
and the convention of summation over repeated indices is understood.
If $(p_{\mu})$ are coordinates on the fibers of the cotangent bundle
$T^*\manX$, the operators $x^{\mu}$ and $p_{\mu}$ and their
commutation relations~$\commut{p_{\nu}}{x^{\mu}}
=-\imath\delta^{\mu}_{\nu}$ are represented on sections
$\Gamma(\bundle{\cH})$ of $\bundle{\cH}$ in the standard way through
multiplication and derivation with~$x^{\mu}$.

We assume that under quantization, the BRST charge $\brst^{cl}$
becomes a nilpotent Hermitian operator~$\brst$ on
$\Gamma(\bundle{\cH})$.  We additionally require~$\brst$ to contain a
finite number of derivatives with respect to $x$.
The first-quantized system $(\brst,{\Gamma(\bundle{\cH}}))$ is thus
determined by the BRST operator $\brst$ and the space of states
$\Gamma({\bundle{\cH}})$.

\subsection{Local field theory and master action}\label{sec:LFT}
We now associate a local field theory with the first-quantized system
$(\brst,{\Gamma(\bundle{\cH}}))$.  Given a real basis $(e_A)$ in
$\cH$, we consider the dual basis $(\psi^{A})$, with the assignments
$\gh{\psi^A}=-\gh{e_A}$ and with the Grassmann parities
$\p{\psi^A}=\p{e_A}$, and let $\assalgA_{\cH}$ be the algebra of
polynomials in the $(\psi^{A})$.  We then consider $(\psi^{A})$ as
coordinates on a supermanifold $\manM_\cH$ (that is, $\manM_\cH$ is
the affine supermanifold whose algebra of functions is the associative
supercommutative algebra $\assalgA_{\cH}$).  We view $\manM_\cH$ as a
real supermanifold.  The inner product
$\inner{{}\cdot{}}{{}\cdot{}}_\cH$ on $\cH$ then determines two forms
on $\manM_\cH$, a symmetric and a symplectic one, in accordance with
(see \cite{Kibble:1979tm,Heslot:1985xx} and also
\cite{Hatfield:1992rz,Schilling:1996xx,Ashtekar:1997ud,
  Barnich:2003wj})\footnote{In the string field theory literature, a
  complex bilinear version of the symplectic structure is considered
  as a starting point and is related to the sesquilinear inner product
  by an appropriate antilinear involution (see,
  e.g.,~\cite{Gaberdiel:1997ia}).  On the associated real form of
  $\cH$, both these symplectic structures coincide.}
\begin{equation}\label{decomp-inner}
  \begin{gathered}
    \inner{\psi}{\phi}_\cH=g(\psi,\phi)+i\omega(\psi,\phi),\\
    g_{AB}=(-1)^{(\kappa+1)\p{e_A}}g(e_{A},e_{B}), \qquad
    \omega_{AB}=(-1)^{(\kappa+1)\p{e_A}}\omega(e_A,e_B),
  \end{gathered}
\end{equation}
where we use $\kappa$ to denote the Grassmann parity of
$\inner{{}\cdot{}}{{}\cdot{}}_\cH$ and hence of the forms $g$
and~$\omega$. In what follows, we only consider the case where the
inner product is odd and of the ghost number~$-1$.

We assume that $\cH$ is decomposed with respect to the ghost number,
\begin{equation}\label{H-decomp}
  \cH=\bigoplus_k\cH^{(k)},
\end{equation}
where elements of $\cH^{(k)}$ are of ghost number $-k$.  If $e_{A_k}$
denotes basis elements in $\cH^{(k)}$, then
$\gh{\psi^{A_k}}=k=-\gh{e_{A_k}}$.

We next take the space of fields to be the space of (suitably smooth)
maps from $\manX$ to the supermanifold $\manM_{\cH}$.  The fields
$\psi^{A}(x)$ then inherit the ghost number and Grassmann parity.  In
particular, $\psi^{A_0}(x)$ denote fields associated with basis
elements in the zero-ghost-number subspace $\cH^{(0)}$. These are
interpreted as physical fields; the remaining fields are identified
with auxiliary BRST variables (ghost fields and antifields).

We define the action for the physical fields as
\begin{equation}\label{eq:physaction0}
  \bundle{S}^{\mathrm{ph}}[\psi]
  =-\half\int  d^dx\,
  g_{A_0B_{-1}}\psi^{A_0}\,{\brst}^{B_{-1}}_{C_0} \,\psi^{C_0},
\end{equation}
where the differential operators ${\brst}^{B_{-1}}_{C_0}$ are defined
as
\begin{equation}
  \begin{gathered}
    \brst \bundle\phi=\brst (e_{A_0}\bundle\phi^{A_0}(x))
    =e_{B_{-1}}\brst^{B_{-1}}_{A_0}\bundle\phi^{A_0}(x), \quad
    \bundle\phi\in\Gamma(\bundle\cH^{(0)})
  \end{gathered}
\end{equation}
and $\Gamma(\bundle\cH^{(0)})$ is the space of zero-ghost-number
sections of $\bundle\cH$.  The Euler--Lagrange equations of motion
following from $S^{\mathrm{ph}}$ are given by
\begin{equation}\label{eq:eom}
  \brst^{A_{-1}}_{B_0} \psi^{B_0}(x)=0.
\end{equation}
Because $\brst^2=0$, the action~\eqref{eq:physaction0} possesses the
gauge symmetries $\delta_\epsilon
\psi^{A_0}={\brst^{A_0}}_{B_1}\epsilon^{B_1}$, for some gauge
parameters $\epsilon^{B_1}$. For the same reason, these gauge
symmetries are in general reducible.

Before giving the Batalin--Vilkovisky master action associated with
this gauge theory, we introduce the concept of \textit{string field}
in the context of local field theory. Let $\bundle{\assalgA}$ be the
\textit{algebra of local functions} associated with the fields
$\psi^{A}(x)$, i.e., the algebra of functions depending on $x^\mu$,
$\psi^A$, and $\psi^A_{\mu_1}$, $\psi^A_{\mu_1\mu_2}$, \dots,
$\psi^A_{\mu_1\dots\mu_n}$ up to some finite order (see,
e.g.,~\cite{Olver:1993}, \cite{Andersonbook}). The ghost-number
assignments and the Grassmann parities of $\psi^A_{\mu_1\dots\mu_n}$
coincide with those of $\psi^A$.  The space of local functionals
$\cF_{\bundle{\assalgA}}$ is the quotient of the space of local
functions modulo those of the form $\dd_\mu j^\mu$, where
$\partial_{\mu}$ denotes the total derivative with respect to $x^\mu$,
i.e., the vector field
\begin{equation}\label{d-total}
  \dd_\mu=\dl{x^\mu}+\psi^A_{,\mu}\dl{\psi^A}+
  \psi^A_{,\mu\nu}\dl{\psi^A_{,\nu}}+\ldots.
\end{equation}
We use the integral sign $\int d^\dmn x$ to denote projection from
local functions to local functionals.

We consider the right $\bundle{\assalgA}$-module
$\cH\tensor\bundle{\assalgA}$.  The sesquilinear form on $\cH$ is
naturally extended to $\cH\tensor\bundle\assalgA$ as
\begin{equation}\label{iyt:2-5}
  \inner{\psi\tensor \bundle{f}}{\phi\tensor \bundle{g}}
  =(-1)^{\p{\bundle{f}}\p{\bundle{\phi}}}
  \inner{\psi}{\phi}_\cH\,\bundle{f}\bundle{g},\quad
  \bundle{f},\bundle{g}\in \bundle\assalgA,
\end{equation}
and the forms $g(\cdot,\cdot)$ and $\omega(\cdot,\cdot)$
(see~\eqref{decomp-inner}) are extended similarly.  Applying the
projection $\int d^\dmn x$ to~\eqref{iyt:2-5} gives a bilinear form on
$\cH\tensor\bundle{\assalgA}$ with values in local functionals.

The odd symplectic structure on $\manM_\cH$ induces a map
$(\cdot,\cdot): \cF_{\bundle{\assalgA}}\otimes
\bundle{\assalgA}\longrightarrow \bundle{\assalgA}$ defined by
\begin{equation}
  \label{eq:1}
  (\bundle{F},\bundle{g})=\vddr{\bundle{f}}{\psi^A}
  \omega^{AB}\ddl{\psi^B}{\bundle{g}}+\partial_\mu(
  \vddr{\bundle{f}}{\psi^A}\omega^{AB})\ddl{\psi^B_{,\mu}}{\bundle{g}}
  +\dots,
\end{equation}
where $\bundle{F}=\int d^\dmn\ x {\bundle{f}}\in
\cF_{\bundle{\assalgA}}$ and ${\bundle{g}}\in \bundle{\assalgA}$ with
$\vddl{}{\psi^A}$ denoting the Euler--Lagrange derivative with respect
to $\psi^A$.  We note that the vector field $({\bundle{F}},\cdot)$
commutes with $\partial_\mu$, which implies that there is a
well-defined bracket induced in the space of local functionals. This
bracket can be shown to be an odd graded Lie bracket, called
antibracket in what follows.

Linear operators acting on $\Gamma(\bundle\cH)$ are also extended to
$\cH\tensor\bundle{\assalgA}$.  Namely, identifying the space of
linear operators on~$\Gamma(\bundle\cH)$ with ${\rm
  Diff}(\manX)\tensor\End(\cH)$, where ${\rm Diff}(\manX)$ is the
space of differential operators on $\manX$ (and $\End(\cH)$ is the
space of linear operators on $\cH$), we define
\begin{equation}
  (A\tensor O)(\phi\tensor\bundle{f})=(A\phi)\tensor(O\bundle f), \quad
  A\in \End(\cH),\,
  O\in {\rm Diff}(\manX), \,
  \phi\in\cH, \,
  \bundle{f}\in \bundle{\assalgA},
\end{equation}
where the action of $O$ on $\bundle{f}\in \bundle{\assalgA}$ is
defined by replacing each $x^\mu$-derivative with~$\dd_\mu$ given
in~\eqref{d-total}.

The \textit{string field} $\Psi$ is the element of the module
$\cH\tensor\bundle{\assalgA}$ defined as
\begin{equation}\label{eq:SF}
  \Psi=e_A\tensor\psi^A.
\end{equation}
We omit $\tensor$ in what follows. We often refer to $\Psi$ as the
\textit{string field associated with~$\cH$}.  For any subspace
$\cV\subset\cH$, there is a restriction of $\Psi$ to a string field
associated with~$\cV$.  In particular, taking $\cV$ to range over the
subspaces $\cH^{(k)}$ in~\eqref{H-decomp} gives the decomposition
$\Psi=\sum_k\Psi^{(k)}$ with
$\Psi^{(k)}=e_{A_k}\psi^{A_k}=\Psi\big|_{\cH^{(k)}}$; the components
of $\Psi^{(0)}$ are physical fields $\psi^{A_0}$.  In terms of the
string field, expression~\eqref{eq:physaction0} takes the form
\begin{equation}\label{eq:physaction}
  \bundle{S}^{\mathrm{ph}}[\Psi^{(0)}]=-\half\int  d^dx\,
  \inner{\Psi^{(0)}}{\brst \Psi^{(0)}}.
\end{equation}

The local functional
\begin{equation}\label{eq:psiopsi}
  \bundle{S}[\Psi]=-\half\int  d^{\dmn}x\ \inner{\Psi}{\brst \Psi},
\end{equation}
satisfies the master equation $\ab{\bundle{S}}{\bundle{S}}=0$ and
$\bundle{S}[\Psi]\big|_{\Psi^{(l)}=0,l\neq
  0}=\bundle{S}^{ph}[\Psi^{(0)}]$ and is therefore the master
action~\cite{Thorn:1987qj,Bochicchio:1987zj,Bochicchio:1987bd,
  Thorn:1989hm} associated with~\eqref{eq:physaction}. We then
consider the BRST differential
\begin{equation}\label{sS}
  s=\ab{\bundle{S}}{{}\cdot{}}.
\end{equation}
The cohomology of this differential in the space of local functionals
contains information on consistent interactions and on (generalized)
global symmetries~\cite{Barnich:1993vg,Barnich:1995db,Barnich:2000zw}.

\subsubsection*{Remarks}\mbox{}

(i)~Alternatively, one can think of $\Gamma(\bundle{\cH})$ as a
Hilbert space with basis vectors $e_A^x=|x\rangle\tensor e_A$.
Instead of \eqref{eq:SF}, one then obtains
\begin{equation}
  \label{eq:xSF}
  {\mathbf\Psi}=\int dx\,\, |x\rangle\tensor e_A\, \,\psi^A(x)
\end{equation}
as the string field.  We prefer using $\Psi$ defined in~\eqref{eq:SF}
in order to avoid additional assumptions needed to work
with~\eqref{eq:xSF}.

\medskip

(ii)~In constructing the field theory, we used real fields $\psi^A$
associated with a real basis $(e_A)$ in $\cH$.  Equivalently, one can
use complex fields associated with a complex basis in $\cH$ (see,
e.g.,~\cite{Gaberdiel:1997ia,Barnich:2003wj} for more details).  This
option is used in Secs.~\bref{sec:Fronsdal} and~\bref{sec:KG}.

\subsection{Generalized auxiliary fields in the Lagrangian
  context}\label{sec:auxiliary}
For a Lagrangian field theory determined by
$\bundle{S}^{\mathrm{ph}}$, \textit{ordinary auxiliary fields} are
fields $v^a$ whose Euler--Lagrange equations
$\vddl{\bundle{S}^{\mathrm{ph}}}{v^a}=0$ can be solved algebraically
for $v^a$. In this subsection we recall the definition of
\textit{generalized auxiliary fields} in the Lagrangian context
\cite{Dresse:1990dj}.

For a given action $\bundle{S}[\psi]$ that depends on \textit{some}
variables $\psi^A$ and satisfies the master equation
$(\bundle{S},\bundle{S})=0$, we suppose that the $\psi^A$ can be split
into $w^a,w^*_a,\varphi^\gamma$ such that the equations
\begin{equation}\label{eq:var-S}
  \vddl{\bundle{S}}{w^a}=0
\end{equation}
can be solved algebraically for $w^a$ at $w^*_a=0$.  Let then
$w^a=W^a[\varphi]$ be the solution, i.e.,
\begin{equation}
  \vddl{\bundle{S}}{w^a}\Bigm|_\Sigma\equiv0,
\end{equation}
where $\Sigma$ is determined by the constraints
$w^a-W^a[\varphi]=0,~w^*_a=0$ and their space-time derivatives. If
$w^a$ and $w^*_a$ are canonically conjugate with respect to the
antibracket (and therefore the defining relations for $\Sigma$ are
second-class constraints), we say that $w^a,w^*_a$ are generalized
auxiliary fields.

Generalized auxiliary fields can be consistently eliminated in the
sense that the master action pulled back to $\Sigma$ satisfies the
master equation with respect to the corresponding Dirac antibracket.
Elimination of generalized auxiliary fields is a natural equivalence
of local gauge theories, in particular, (local) BRST cohomology groups
are invariant under the elimination of generalized auxiliary fields.
In addition to the elimination of ordinary auxiliary fields and their
conjugate antifields, the elimination of generalized auxiliary fields
includes the elimination of pure (algebraic) gauge degrees of freedom
together with their associated ghosts and antifields.

We now give an example of generalized auxiliary fields which can
already be identified at the first-quantized level (see
\cite{Barnich:2003wj} for more details).  Consider the situation where
$\cH$ has the form $\cH =\cU\otimes \cV$ and where the inner product
on $\cH$ is the tensor product of an inner product on $\cU$ and an
inner product $\inner{{}\cdot{}}{{}\cdot{}}_\cV$ on $\cV$.  Suppose
that the BRST differential on $\Gamma(\bundle{\cH})\simeq
\Gamma(\bundle{\cU})\otimes \cV$ is of the form
$\brst=\brst_\cU\otimes\mathbf{1}+\mathbf{1}\otimes \brst^\prime$,
where $\brst^\prime$ is a nilpotent Hermitian operator on $\cV$ with
respect to $\inner{\cdot}{\cdot}_\cV$.  Suppose furthermore that $\cV$
decomposes into a singlet and null doublets as well as quartets with
respect to $\brst^\prime$ and $\inner{{}\cdot{}}{{}\cdot{}}_\cV$ (see,
e.g.,~\cite{Henneaux:1992ig} for details on irreducible
representations of BRST algebra).  Then the fields carrying the index
associated with the null doublets are ordinary auxiliary fields and
their associated antifields, while those carrying the index of
quartets are generalized auxiliary fields.  After their elimination,
we obtain an action of the form~\eqref{eq:psiopsi} associated
with~$\cU$~alone.

\section{Non-Lagrangian BRST field theory}
\label{sec:NL}
In the previous section, we constructed a particular Lagrangian gauge
field theory. We now show that some basic features of the construction
can be extended to the non-Lagrangian context.  For a BRST
first-quantized system, we just take over the previous definitions of
equations of motion and of gauge symmetries.  For possibly nonlinear
gauge field theories defined by a differential $s$, we define
generalized auxiliary fields in terms of the differential $s$ and show
that their elimination or introduction does not affect various
cohomologies of $s$.  We then go back to the not necessarily
Lagrangian field theory defined by a BRST first-quantized system and
study the first-quantized counterpart of generalized auxiliary fields
and their elimination. We show that the existence of a Lagrangian for
the field theory associated with a BRST first-quantized system is
equivalent to the existence of an inner product that makes the BRST
operator formally self-adjoint.  Finally, we illustrate these concepts
in the case of the Fronsdal system.

\subsection{Equations of motion and gauge symmetries from a BRST
  differential}\label{sec3.1}
In the case where an inner product does not exist on $\cH$ (or is not
specified), one can still consider the equations of motion
\eqref{eq:eom} for the physical fields
$\psi^{{A_0}},\,\gh{\psi^{{A_0}}}=0$ contained in $\Psi^{(0)}$.  These
equations are then not necessarily equivalent to variational ones. In
this context, one can also define the BRST differential
\begin{equation}\label{brstwj}
  s=(\brst\Psi)^A\dl{\psi^A}
  +\partial_{\mu}[(\brst\Psi)^A]\dl{\psi^A_{,{\mu}}}
  +\dots,
\end{equation}
satisfying $\commut{s}{\partial_{\mu}}=0$ where $\partial_\mu$ is
defined in~\eqref{d-total}. The BRST differential determines the
equations of motion through
\begin{gather}
  s\Psi^{(-1)}\equiv \brst\Psi^{(0)}=0.\label{eqofmo}
\end{gather}

The differential $s$ in~\eqref{brstwj} is linear in~$\Psi$.  More
generally, we consider a not necessarily linear differential (still
denoted by $s$ and called the BRST differential) acting on a space of
local functions.  We write $\psi^A(x)$ for the components of these
functions with respect to a basis~$(e_A)$, with $\Psi(x)=
e_A\psi^A(x)$.  As before, we also assume that the space of fields is
graded by ghost number, see~\eqref{H-decomp}, with $\Psi^{(\ell)}$
being the string field associated with $\cH^{(\ell)}$.  The BRST
differential $s$ is considered to have ghost number $1$.  This
differential determines equations of motion for the physical fields~as
\begin{equation}\label{nonlineom}
  s\Psi^{(-1)}|_{\Psi^{(\ell)}=0,\,\ell\neq 0}=0,
\end{equation}
By expanding $s^2=0$ according to the ghost number, it can be shown
that the transformations
\begin{equation}
  \delta\Psi^{(0)}=s\Psi^{(0)}|_{\Psi^{(\ell)}=0,\,\ell\neq 0,1},
\end{equation}
with ghost-number-$1$ component fields of $\Psi^{(1)}$ replaced by
gauge parameters, are gauge symmetries of
Eqs.~\eqref{nonlineom}.\footnote{By gauge symmetries of equations of
  motion~\eqref{nonlineom}, we understand transformations that depend
  on arbitrary parameters and that leave the ideal of local functions
  generated by the left-hand sides of~\eqref{nonlineom} invariant.}

\subsection{Generalized auxiliary fields in the non-Lagrangian
  context}\label{subsec:nlgaf}
Generalized auxiliary fields, defined in~\bref{sec:auxiliary} in the
Lagrangian context, can also be defined for a non-Lagrangian system
described by some differential~$s$ that is not necessarily linear
in~$\psi^A$.

Let $\cV$ be an even-dimensional subspace in $\cH$ and let
$(e_{\alpha}, f_a, g_a)$, be a basis in~$\cH$ such that $(f_a,g_a)$
form a basis in $\cV$.  Let then $\varphi^{\alpha}$, $v^a$, $w^a$ be
the fields associated with the respective basis elements $e_\alpha$,
$f_a$, $g_a$.  We also suppose that the equations
\begin{equation}\label{eq:swa}
  (sw^a)|_{w^a=0}=0
\end{equation}
can be solved algebraically for $v^a$ as $v^a=V^a[\varphi]$.  In this
context, we still refer to the fields $v^a$ and $w^a$ as
\textit{generalized auxiliary fields}.

The differential $s$ then naturally restricts to a differential
$\tilde s$ on the surface $\Sigma$ defined by
\begin{equation} \label{eq:constraints}
  w^a=0,\qquad v^a-V^a[\varphi]=0,
\end{equation}
and their space-time derivatives. Indeed, $s$ preserves the ideal of
local functions vanishing on $\Sigma$: because this ideal is generated
by $w^a$, $v^a-V^a[\varphi]$, and their derivatives with respect to
coordinates $x^{\mu}$ on the base manifold, it suffices to show that
$sw^a|_\Sigma=0$ and $s(v^a-V^a[\varphi])|_\Sigma=0$.  The first
equation is tautological, while the second is a consequence of the
nilpotency of $s$ and the representation
\begin{equation}
  v^a-V^a[\varphi]=A^{a}_{b} sw^b+B^{a}_{b} w^b,
\end{equation}
where $A^a_b$ and $B^a_b$ are differential operators and standard
regularity conditions on Eqs.~\eqref{eq:constraints} have been
assumed.

It follows from the definition that generalized auxiliary fields can
be organized in subsets $v^{a_{k}}$ and $w^{a_{k-1}}$ of the
respective ghost numbers $k$ and $k-1$

\begin{prop}\label{prop:aux}
  Let $v^a$ and $w^a$ be generalized auxiliary fields.  Let also
  $\gh{v^a}=k$ and $\gh{w^a}=k-1$.

  \begin{enumerate}
    
  \item\label{1} For $k\neq 0,1$, Eqs.~\eqref{nonlineom} are unchanged
    by elimination of~$v,w$.
    
  \item\label{2} For $k=0$, the fields $v^a$ are ordinary auxiliary
    fields in the sense that Eqs.~\eqref{nonlineom} can be solved
    algebraically for $v^a$.
    
  \item\label{3} For $k=1$, the fields $w^a$ are pure gauge in the
    sense that they are not constrained by Eqs.~\eqref{nonlineom}.
    Furthermore, arbitrary shifts in these variables can be generated
    by gauge transformations determined by $s$ for an appropriate
    choice of gauge parameters.

  \end{enumerate}
\end{prop}
\begin{proof}
  Assertion~\ref{1} follows because for $k,k-1\neq 0$, the fields
  $v^a,w^a$ have nonvanishing ghost numbers, and therefore do not
  enter the equations of motion.
  
  For $k=0$, the fields $v^a$ carry ghost number $0$ and the equations
  $sw^a|_{w=0}=0$ can be solved algebraically for $v^a$.  This implies
  that the equations $sw^a|_{\Psi^{(\ell)}=0,\,\ell\neq 0}=0$ can
  still be solved algebraically because no $v^a$ are set to zero by
  putting all $\Psi^{(l)}$ for $l\neq 0$ to zero.  At the same time,
  these equations form a subset of Eqs.~\eqref{nonlineom}, and we
  conclude that $v^a$ are ordinary auxiliary fields.
  
  To show assertion~\ref{3}, we introduce new independent fields
  $w^a$, $\chi^a=sw^a$, and $\varphi^{\alpha}$.  In these coordinates,
  $s$ takes the form
  \begin{equation}\label{eq:useful}
    s=Q^{\alpha}\dl{\varphi^{\alpha}}+\chi_a\dl{w^a}
    +\partial_{\mu} Q^{\alpha}\dl{\varphi^{\alpha}_{,{\mu}}}
    +\partial_{\mu}\chi_a\dl{w^a_{,{\mu}}}
    +\dots,
  \end{equation}
  where $Q^{\alpha}=s\varphi^{\alpha}$, which implies the statement
  about shift symmetries.  We can further modify this coordinate
  system by redefining $\varphi^{\alpha}$ to $\varphi^{1
    \alpha}=\varphi^\alpha-\rho Q^\alpha$, where
  $\rho=w^a\frac{\partial}{\partial \chi^a}$.  Then $s\varphi^{1
    \alpha}\equiv Q^{1 \alpha}$ can be easily seen not to contain
  terms linear in $\chi^a$.  {}From the consistency condition
  $sQ^{1\alpha}=0$, it then follows that $Q^{1 \alpha}$ does not
  contain terms linear in $w^a$ either.  Induction on homogeneity
  degree in both $\chi^a,w^a$ together with expansions in $\chi^a$
  then shows, along similar lines, that the dependence on $\chi^a$ and
  $w^a$ can be completely absorbed through all orders by appropriate
  redefinitions of $\varphi^\alpha$.  This implies that the equations
  of motion $Q^{\alpha}|_{\Psi^{(\ell)}=0,\,\ell\neq 0}=0$ do not
  involve $w^a$ in these coordinates.
\end{proof}

Given the BRST differential $s$, we can consider its cohomology groups
in the space of local functions, in the space of local functionals,
the cohomology of the adjoint action \hbox{$\commut{s}{{}\cdot{}}$} in
the space of evolutionary vector fields or the cohomology of $s$
modulo $\derham=dx^{\mu}\dd_{\mu}$ in the space of horizontal forms.
Here, evolutionary vector fields are those which commute with
$\dd_{\mu}$ (see, e.g., \cite{Olver:1993}), horizontal forms are
differential forms in $dx^{\mu}$ with coefficients in local functions,
and $\dd_{\mu}$ is the total derivative. As a consequence of
representation~\eqref{eq:useful}, we see that all the above cohomology
groups are invariant under elimination of generalized auxiliary
fields.  Somewhat more formally, we summarize this as the following
proposition.
\begin{prop}
  If $v^a$ and $w^a$ are generalized auxiliary fields for a
  differential $s$ and $\tilde s$ is the restriction of~$s$ to the
  surface $\Sigma$ defined by~\eqref{eq:constraints}, then $s$ and
  $\tilde s$ have the same cohomology.
\end{prop}

\subsection{Elimination of generalized auxiliary fields at the
  first-quantized level}
\label{sec:elimination-1st-q}
We now concentrate on the case where the BRST differential $s$
originates from the first-quantized BRST operator $\brst$ as
in~\eqref{brstwj}.  At the first-quantized level, we can then identify
the states that correspond to (generalized) auxiliary fields.  Namely,
as a first-quantized counterpart of the setting at the beginning of
Sec.~\bref{subsec:nlgaf}, we suppose that
\begin{equation}\label{eq:EFG-decomp}
  \cH=\cE\oplus\cF\oplus \cG,\qquad
  \dim\cG=\dim\cF
\end{equation}
(where in the graded-finite-dimensional case, the dimensions of the
corresponding graded components must be equal).  Elements of
$\Gamma(\bundle{\cH})$ and the BRST operator can then be decomposed
accordingly,
\begin{equation}
  \begin{gathered}
    \bundle{\phi}(x)=\bundle{\phi}^\cE(x)+\bundle{\phi}^\cF(x)
    +\bundle{\phi}^\cG(x),\\
    \st{\bullet_1\bullet_2}\brst \bundle{\phi}|_{\bullet_2}=(\brst
    (\bundle{\phi}|_{\bullet_2}))|_{\bullet_1},
  \end{gathered}
\end{equation}
where $\bullet_1,\bullet_2$ run over $\cE,\cF,\cG$.  Of course, this
decomposition extends to the string field:
$\Psi=\Psi^\cE+\Psi^\cF+\Psi^\cG$, where $\Psi^\cX=\Psi\bigm|_\cX$
(see Sec.~\bref{sec:LFT}) for simplicity of notation.

We note that here and below, we consider differential operators with a
finite (or graded finite) number of derivatives. If an operator is
invertible in this space, we call it \textit{algebraically}
invertible.

\begin{lemma}\label{deffirst}
  Given decomposition~\eqref{eq:EFG-decomp}, the fields $v^a$ and
  $w^a$ associated with the respective basis elements $f_a$ and $g_a$
  of $\cF$ and $\cG$ are generalized auxiliary fields if and only if
  $\st{\cG\cF}\brst$ is algebraically invertible.
\end{lemma}

\begin{proof}
  Indeed, the equations $(sw^a)|_{w=0}=0$ become $((\brst
  \Psi)|_{\cG})|_{\Psi^\cG=0}=0$, or equivalently,
  \begin{equation}\label{aux1}
    \st{\cG\cF}{\brst}\Psi^\cF
    +\st{\cG\cE}{\brst}\Psi^\cE=0.
  \end{equation}
  This equation can be solved algebraically for the component fields
  of $\Psi^\cF$ in terms of the component fields of $\Psi^\cE$ and a
  (graded) finite number of derivatives if and only if
  $\st{\cG\cF}\brst$ is invertible as a differential operator with the
  inverse containing a (graded) finite number of derivatives. Hence,
  $\cF\oplus \cG\subset \cH$ corresponds to $\cV$ in the definition of
  generalized auxiliary fields in \bref{subsec:nlgaf}.
\end{proof}

The reduced differential $\tilde s$ is determined by the BRST operator
\begin{equation}\label{tilde-brst}
  \tilde\brst
  =(\st{\cE\cE}{\brst}
  - \st{\cE\cF}{\brst}(\st{\cG\cF}{\brst})^{-1}
  \st{\cG\cE}{\brst})
  :\Gamma(\bundle{\cE})\to\Gamma(\bundle{\cE})
\end{equation}
such that
\begin{equation}\label{tildebrst}
  \tilde s\Psi^\cE
  \equiv(\brst\Psi)^\cE|_{\Psi^\cG=0,
    \Psi^\cF=\Psi^\cF(\Psi^\cE)}
  = {\tilde\brst}\Psi^\cE,
\end{equation}
where we write $\Psi^\cF=\Psi^\cF(\Psi^\cE)$ for the solution
of~\eqref{aux1} for~$\Psi^\cF$. Hence,
\begin{prop}\label{prop:add} Let $\st{\cG\cF}\brst$ be
  algebraically invertible. By elimination of the corresponding
  generalized auxiliary fields, the field theory associated with
  $(\brst,\Gamma(\bundle{\cH}))$ is reduced to the field theory
  associated with $(\tilde\brst,\Gamma(\bundle{\cE}))$, where
  $\tilde\brst$ is defined in \eqref{tilde-brst}.
\end{prop}

We call the reduction from the system $(\brst,\Gamma(\bundle{\cH}))$
to $(\tilde\brst,\Gamma(\bundle{\cE}))$ \textit{algebraic} because in
the process of reduction, we stay in the space of differential
operators with graded finite number of derivatives and because at the
field theory level, this reduction corresponds to the algebraic
elimination of generalized auxiliary fields.

\begin{prop}\label{prop:equivalent}
  Let $\st{\cG\cF}\brst$ be algebraically invertible and $\tilde\brst$
  be defined by~\eqref{tilde-brst}. The first-quantized systems
  $(\brst,\Gamma(\bundle{\cH}))$ and
  $(\tilde\brst,\Gamma(\bundle{\cE}))$ are then equivalent in the
  sense that $H(\brst,\Gamma(\bundle{\cH}))\simeq
  H(\tilde\brst,\Gamma(\bundle{\cE}))$.
\end{prop}
\begin{proof}
  Indeed, as we see momentarily, the map
  \begin{equation} \label{eq:imap}
    \begin{split}
      \incmap : \Gamma(\bundle{\cE})&\rightarrow
      \Gamma(\bundle{\cH}),\\[-4pt]
      \bundle{\phi}^\cE(x)&\mapsto
      \bundle{\phi}^\cE(x)-(\st{\cG\cF}{\brst})^{-1}\st{\cG\cE}{\brst}
      \bundle{\phi}^\cE(x)
    \end{split}
  \end{equation}
  is a morphism of complexes
  $(\tilde\brst,\Gamma(\bundle{\cE}))\to(\brst,\Gamma(\bundle{\cH}))$.
  We note that $\incmap$ has no image in~$\Gamma(\bundle{\cG})$, while
  $(\st{\cG\cF}{\brst})^{-1}\st{\cG\cE}{\brst}
  \bundle{\phi}^\cE(x)\in\Gamma(\bundle{\cF})$.  That $\incmap\,
  \tilde\brst=\brst\incmap$ follows from the identity
  \begin{equation}\label{id1}
    \st{\cF\cE}{\brst}
    -\st{\cF\cF}{\brst}(\st{\cG\cF}{\brst})^{-1}\st{\cG\cE}{\brst}
    =- (\st{\cG\cF}{\brst})^{-1}\st{\cG\cE}{\brst}\tilde\brst,
  \end{equation}
  which in turn can be established by straightforward algebra using
  the nilpotency of $\brst$ in the different subspaces.  Furthermore,
  the map $\tilde\incmap$ induced in cohomology can easily be shown to
  be an isomorphism.
\end{proof}

One possibility to find a decomposition $\cH=\cE\oplus\cF\oplus\cG$
with $\smash{\st{\cG\cF}{\brst}}$ invertible is as follows.  We
suppose $\cH$ to be equipped with an additional grading besides the
ghost number,
\begin{equation}
  \cH=\bigoplus_{i\geq 0} \cH_{i},\qquad \deg(\cH_{i})=i,
\end{equation}
extend the degree from $\cH$ to $\cH$-valued sections, and let the
BRST operator $\brst$ have the form
\begin{equation} \label{eq:2diffeq}
  \brst= \brst_{-1}
  +\brst_0
  +\sum_{i\geq1}\brst_i,\qquad \deg(\brst_i)=i,
\end{equation}
with $\brst_i:\Gamma(\bundle{\cH})_{j}\to\Gamma(\bundle{\cH})_{i+j}$.
\footnote{In the case where equations of motion have the form
    $\Omega\Phi=0$ with $\Phi$ a collection of fields which are
    differential $p$-forms and $\Omega$ a flat covariant differential,
    the dynamical fields were identified with the cohomology of
    $\Omega_{-1}$ at form degree $p$ already
    in~\cite{Shaynkman:2000ts,Vasiliev:2001zy}.}
\begin{prop}\label{prop:red}
  Let $\brst_{-1}$ be independent of $x$ and $x$-derivatives and let
  $\cH=\cE\oplus\cF\oplus\cG$ with $\Ker \brst_{-1}\supset \cE\simeq
  H(\brst_{-1},\cH)$ and $\cG= {\rm Im}\,\brst_{-1}$.  Then
  $\st{\cG\cF}{\brst}$ is algebraically invertible, and hence the
  system $(\brst,\Gamma(\bundle{\cH}))$ can be algebraically reduced
  to $(\tilde\brst,\Gamma(\bundle{\cE}))$.
\end{prop}
\begin{proof}
  Because $\brst_{-1}$ contains no $x$ dependence and no
  $x$-derivatives and because the space $\cH$ is (graded)
  finite-dimensional, we have decomposition~\eqref{eq:EFG-decomp},
  where $\cE\oplus \cG=\Ker\brst_{-1}$, $\cG=\im\brst_{-1}$, and $\cF$
  is a complementary subspace, which is isomorphic to $\cG$.  We can
  then construct the operator $\rho=(\st{\cG\cF}{\brst}_{-1})^{-1}$.
  Because $\st{\cG\cF}{\brst}=\st{\cG\cF}{\brst}_{-1}+\dots$, it can
  be formally inverted order by order in the degree, with the inverse
  given by
  \begin{equation} \label{eq:2}
    \smash[b]{
      (\st{\cG\cF}{\brst})^{-1}=\sum_{n\geq 0}(-1)^n\rho[(\sum_{i\geq 0}
      \st{\cG\cF}{\brst}_i)\rho]^n.}
  \end{equation}
\end{proof}
If in addition the cohomology of $\brst_{-1}$ is concentrated in one
degree, then $\tilde\brst$ is the BRST operator induced by $\brst_0$
in~$\cE$, ${\tilde\brst}=\st{\cE\cE}{\brst_0}$.  This follows because
under the assumptions of Proposition \bref{prop:red}, the degree of
$(\st{\cG\cF}{\brst})^{-1}$ is strictly positive, and hence so is the
degree of the second term in the definition of $\tilde\brst$
in~\eqref{tilde-brst}, which therefore cannot contribute.

\medskip

In fact, the invertibility of $\st{\cG\cF}{\brst}$ implies the
existence of contractible pairs of the form
\begin{equation}\label{3.18}
  \brst \tilde e_\alpha= \tilde e_\beta\tilde\brst^\beta_\alpha,
  \qquad
  \brst \tilde f_a=\tilde g_a,
\end{equation}
but in a bigger space, namely the free $\cD$-module generated by
$\cH$.

Indeed, we can consider the complex $(\brst,\cD(\bundle{\cH}))$ with
$\cD({\bundle\cH})=\cH \tensor {\rm Diff} (\manX)$ where ${\rm
  Diff}(\manX)$ is the algebra of differential operators on $\manX$.
$\cD({\bundle\cH})$ is a free ${\rm Diff}(\manX)$-module.  The space
$\Gamma(\bundle{\cH})$ can be naturally identified with a subspace in
$\cD({\bundle\cH})$.  In terms of components, if $(e_A)$ is a basis of
$\cH$, then elements of $\Gamma({\bundle{\cH}})$ are of the form
$\bundle{\phi}(x)=e_A\bundle{\phi}^A(x)$ for some functions
$\bundle{\phi}^A(x)$, while elements of $\cD({\bundle\cH})$ are of the
form $O(x,\dl{x})=e_AO^A(x,\dl{x})$, where $O^A(x,\dl{x})$ are
differential operators.  With respect to $(e_A)$, the action of
$\brst$ in $\cD({\bundle\cH})$ is defined as $\brst
O=e_B\brst^B_AO^A$.  A local frame $(\hat e_A)$ for
$\cD({\bundle\cH})$ is a collection of elements of $\cD({\bundle\cH})$
such that every element $O$ can be uniquely decomposed as $O=\hat
e_A\hat O^A$ with $\hat O^A$ differential operators.  In particular,
the basis $(e_A)$ of $\cH$ induces a local frame of
$\cD({\bundle\cH})$.  Given a frame $(e_A)\equiv (e_\alpha, f_a,
g_a)$, we define a new frame $(\tilde e_A)=(e_BA^B_A)\equiv (\tilde
e_\alpha, \tilde f_a, \tilde g_a)$, with $A^B_A$ invertible
differential operators, through the invertible relations
\begin{gather}
  \tilde e_\alpha= e_\alpha
  -f_b((\st{\cG\cF}{\brst})^{-1}\st{\cG\cE}{\brst})^b_\alpha ,\\
  \tilde f_a= f_b((\st{\cG\cF}{\brst})^{-1})^b_a,\\
  \tilde g_a=
  e_{\beta}(\st{\cE\cF}{\brst}(\st{\cG\cF}{\brst})^{-1})^\beta_a +
  f_b(\st{\cF\cF}{\brst}(\st{\cG\cF}{\brst})^{-1})^b_a+ g_a.
\end{gather}
The expression for $\brst$ in the new frame reduces to the Jordan
form~\eqref{3.18}, as can be checked by using identity~\eqref{id1}.
We call the frame elements $\tilde f_a, \tilde g_a$ algebraically
contractible pairs. The frame $(\tilde e_A)$ can also be used to
expand elements of $\Gamma({\bundle{\cH}})$, $\bundle{\phi}(x)=
e_A\bundle{\phi}^A(x)=\tilde e_A \tilde{\bundle{\phi}}^A(x)$, where
$\tilde e_A$ acts on $\tilde {\bundle{\phi}}^A(x)$.  The components
$\tilde{\bundle{\phi}}^A(x)$ are thus related to the components
$\bundle{\phi}^A(x)$ and their derivatives according to
$\tilde{\bundle{\phi}}^A(x)=(A^{-1})^A_B\bundle{\phi}^B(x)$.  In
Sec.~\bref{sec:LFT}, we have associated fields with (a local frame
induced by) a basis $(e_A)$ of $\cH$.  Here, we are led to take a more
general point of view and associate fields with elements of a local
frame $(\hat e_A)$ in $\cD({\bundle\cH})$.  At the field theory level,
the fields associated with $\hat e_A$ are related to those associated
with $e_A$ through a linear invertible redefinition involving
derivatives.

We have thus shown, by combining with Lemma~\bref{deffirst}:
\begin{prop}
  After an invertible redefinition of the fields and their
  derivatives, generalized auxiliary fields are the fields associated
  with algebraically contractible pairs for $\brst$ in
  $\cD(\bundle{\cH})$.
\end{prop}

\subsection{The action for an equivalence class of equations of motion}
Starting from linear equations of motion~\eqref{eq:eom}, which can be
written as
\begin{equation}
  \brst^{A_{-1}}_{B_0}\psi^{(0)B_0}=0,\label{3.1}
\end{equation}
one can ask the question when are these equations equivalent to
variational ones.

Usually, Eqs.~\eqref{3.1} are defined to be equivalent to variational
ones if there exist field-independent differential operators
\begin{equation}
  \omega_{A_0B_{-1}}
  =\omega^0_{A_0B_{-1}}+\omega^{\mu}_{A_0B_{-1}}\partial_{\mu}
  +\dots,
\end{equation}
with $\omega^0_{A_0B_{-1}}$ invertible, such that
$\omega_{A_0B_{-1}}(\brst^{A_{-1}}_{B_0}\psi^{(0)B_0})$ is given by
Euler--Lagrange derivatives of some quadratic Lagrangian.  Standard
results on the inverse problem of the calculus of variations (see,
e.g.,~\cite{Olver:1993,Anderson1991,Andersonbook}) show that this is
the case if and only if the operator $\brst_{A_0B_0}\equiv
\omega_{A_0C_{-1}}\circ\brst^{C_{-1}}_{B_0}$ is formally self-adjoint.
Here, formal self-adjointness means that
\begin{equation}
  \brst_{A_0B_0}=(-1)^{\p{A_0}\p{B_0}}\brst^{+ C_{-1}}_{A_0}\circ
  \omega^+_{B_0C_{-1}},
\end{equation}
where the adjoint $+$ of an operator
$O=\sum_{k=0}O^{{\mu}_1\dots{\mu}_k}\partial_{{\mu}_1}\dots
\partial_{{\mu}_k}$ is defined as $O^+
f\equiv\sum_{k=0}(-1)^k\partial_{{\mu}_1}\dots
\partial_{{\mu}_k}[O^{{\mu}_1\dots{\mu}_k}f]$ for any local function
$f$.  (We note that the adjoint here is a priori not related to the
conjugation defined on $\cH$.) Whenever such an $\brst_{A_0B_0}$
exists, the action for physical fields can be written as
\begin{equation}\label{3.2}
  \bundle{S}^{\mathrm{ph}}[\psi^{(0)}]=\half\int d^{\dmn}x\
  \psi^{(0)A_0}\brst_{A_0B_0}\psi^{(0)B_0}.
\end{equation}

Similarly, one can ask when the BRST differential $s$
in~\eqref{brstwj} is canonically generated by some quadratic master
action $S$ with respect to some invertible field-independent
antibracket.  This is the case if and only if there exist
field-independent differential operators
\begin{equation}
  \omega_{A_{k}B_{-k-1}}=
  \omega^0_{A_{k}B_{-k-1}}
  +\omega^{\mu}_{A_{k}B_{-k-1}}\partial_{\mu}
  +\dots,
\end{equation}
with $\omega^0_{A_{k}B_{-k-1}}$ invertible and
$\omega_{A_{k}B_{-k-1}}=
(-1)^{1+\p{A_k}\p{B_{-k-1}}}\omega^+_{B_{-k-1}A_k}$, such that the
differential operator $\brst_{A_{k}B_{-k}}\equiv
\omega_{A_{k}C_{-k-1}}\circ\brst^{C_{-k-1}}_{B_{-k}}$ is formally
self-adjoint.  In this case, the master action is given by
\begin{equation}
  \bundle{S}[\psi]=\half\int d^{\dmn}x
  \sum_{k}\psi^{A_{k}}_k\brst_{A_{k}B_{-k}}\psi^{B_{-k}}_{-k},
\end{equation}
while the antibracket is determined through
$(\psi^{A_k}(x),\psi^{B_{-k-1}}(y))=\omega^{A_kB_{-k-1}}\delta(x,y)$.
Furthermore, if a compatible complex structure exists on $\cH$, the
symplectic structure $\omega_{A_{k}B_{-k-1}}$ determines a Hermitian
inner product on~$\Gamma(\bundle{\cH})$.

We see at this stage that the setting in Sec.~\bref{sec:review} can
actually be generalized in that the Hermitian inner product should be
allowed to contain space-time derivatives.

In Secs.~\bref{subsec:nlgaf} and \bref{sec:elimination-1st-q}, we have
discussed when two linear non-Lagrangian theories determined by the
corresponding BRST operators are related via elimination of
generalized auxiliary fields.  Of course, such theories must be
considered equivalent.  It is therefore natural to address the problem
of the existence of a variational principle with this more general
notion of equivalence taken into account, i.e., to consider
Eqs.~\eqref{3.1} equivalent to variational ones if, after elimination
or introduction of generalized auxiliary fields, they are equivalent
to variational ones in the sense defined above.  Similarly, a theory
defined by a BRST differential $s$ should be considered as equivalent
to one with a canonically generated $s$ if, after elimination or
introduction of generalized auxiliary fields, there exists a
symplectic form making the BRST operator formally self-adjoint.

At the first-quantized level, the situation where only one of two
equivalent theories admits a Lagrangian manifests itself in the
existence of two equivalent BRST first-quantized systems
$(\brst,\Gamma(\bundle{\cH}))$ and
$(\tilde\brst,\Gamma(\bundle{\tilde\cH}))$ such that a Hermitian inner
product on the entire space making the BRST operator Hermitian is
known only for one of them.  For instance, suppose that we are in the
setting of Proposition~\bref{prop:red}, where
$\tilde\cH=\cE=H(\brst_{-1},\cH)$, and assume $\tilde\cH$ to be
equipped with an inner product that makes $\tilde\brst$ Hermitian.  In
this case, we can conclude directly at the first-quantized level that
the equations $\brst\Psi^{(0)}=0$ are equivalent to Lagrangian
equations $\tilde\brst\tilde\Psi^{(0)}=0$ via elimination of
generalized auxiliary fields.  Examples of this type occur in the
following sections.

\subsection{First-quantized description of the Fronsdal Lagrangian}
\label{sec:Fronsdal}
As an application of our approach, we show how a simple
first-quantized BRST system gives rise to the sum of the Fronsdal
Lagrangians for free massless higher-spin
fields~\cite{Fronsdal:1978rb}.  The system with which we start was
proposed in~\cite{Ouvry:1986dv,Bengtsson:1986ys} (see
also~\cite{Henneaux:1987cp}).

The variables are $x^\mu,p_\mu,a^\mu,a^{\dagger\mu}$, where
$\mu=1,\dots,\dmn$.  Classically, $x^\mu$ and $p_\mu$ correspond to
coordinates on $T^*\manX$ and $a^\mu,a^{\dagger\mu}$ correspond to
internal degrees of freedom.  After quantization, they satisfy the
canonical commutation relations
\begin{equation}
  \commut{p_\nu}{x^\mu}=-\imath\delta^\mu_\nu,\qquad
  \commut{a^\mu}{a^{\dagger\nu}}=\eta^{\mu\nu}.
\end{equation}
We assume that $x^\mu,p_\mu$ are Hermitian,
$(x^{\mu})^{\dagger}=x^\mu$, $p_{\mu}^{\dagger}=p_\mu$, while $a^\mu$
and $a^{\dagger\mu}$ are interchanged by Hermitian conjugation.

The constraints of the system are
\begin{equation}
  \cL\equiv\eta^{\mu\nu}p_\mu p_\nu=0,\quad
  \cS\equiv p_\mu a^\mu=0,\quad
  \cS^\dagger\equiv p_\mu a^{\dagger\mu}=0.
\end{equation}
For the ghost pairs $(\theta,\cP)$, $(c^\dagger,b)$, and
$(c,b^\dagger)$ corresponding to each of these constraints, we take
the canonical commutation relations in the form\footnote{We use the
  ``super'' convention that $(ab)^\dagger=(-1)^{\p{a}\p{b}}b^\dagger
  a^\dagger$.}
\begin{equation}
  \commut{\cP}{\theta}=-\imath,\qquad \commut{c}{b^\dagger}=1,\qquad
  \commut{b}{c^\dagger}=-1.
\end{equation}
The ghost-number assignments are
\begin{equation}\label{target-gh-n}
  \gh{\theta}=\gh{c}=\gh{c^\dagger}=1,\qquad
  \gh{\cP}=\gh{b}=\gh{b^\dagger}=-1.
\end{equation}
The BRST operator is then given by
\begin{equation} \label{eq:Fbrst}
  \brst_0=\theta\cL+c^\dagger\cS+\cS^\dagger
  c-\imath\cP c^\dagger c.
\end{equation}
The representation space is given by functions of $x^\mu$ (on which
$p_\mu$ acts as $-i\dl{x^\mu}$) with values in the ``internal space''
$\cH_0$.  The latter is the tensor product of the space
$\cH_{\theta,\cP}$ of functions in $\theta$ (coordinate representation
for $(\theta,\cP)$) and the Fock spaces for $(a^\dagger_\mu,a^\mu)$,
$(c^\dagger,b)$, and $(c,b^\dagger)$ with the vacuum conditions
\begin{equation}\label{target-vac}
  a^\mu\vac=b\vac=c\vac=0.
\end{equation}
The inner product  in the space $\cH_0$ is denoted by 
$\inner{\cdot}{\cdot}$.
It is given by the tensor product of the standard Fock space inner 
product and the inner product $\inner{}{}_{\theta,\cP}$ in 
$\cH_{\theta,\cP}$ determined by $\inner{\theta}{1}_{\theta,\, \cP}=1$, 
where $\theta$ and~$1$ are basis
elements.
%

In accordance with the homogeneity degree in $\theta$, the BRST
operator decomposes as
\begin{equation}
  \brst_{0,-1}=-\imath\cP c^\dagger c,\qquad
  \brst_{0,0}=c^\dagger\cS+\cS^\dagger c,\qquad \brst_{0,1}=\theta\cL.
\end{equation}
The assumptions of Proposition~\bref{prop:red} are then satisfied.  In
the present case, we have that the states with the ghost dependence
$\theta b^\dagger$ and $c^\dagger$ form contractible pairs for
$\brst_{0,-1}$, while all other states are representatives of
cohomology classes.  Furthermore, because $\theta b^\dagger$ carries
ghost number $0$, it follows from Proposition \bref{prop:aux} that the
associated fields are ordinary auxiliary fields.

The ghost-number-zero component of the string field is
\begin{equation} \label{eq:SF-phys}
  \Psi^{(0)}=\Phi-\imath\theta b^\dagger \FQ
  +c^\dagger b^\dagger \FR,
\end{equation}
where
\begin{align}
  \Phi&=\sum_{s=0}^\infty \, \ffrac{1}{s!}  a^{\dagger\mu_1}\ldots
  a^{\dagger\mu_s}\vac \,
  \overset{s}{\varphi}_{\mu_1\ldots\mu_s},\notag\\
  \FQ&=\sum_{s=1}^\infty \, \ffrac{1}{(s-1)!}  a^{\dagger\mu_1}\ldots
  a^{\dagger\mu_{s-1}}\vac \, \overset{s}{\fq}_{\mu_1\ldots\mu_{s-1}},
  \label{Phi-C-D}\\
  \FR&=\sum_{s=2}^\infty \, \ffrac{1}{(s-2)!}  a^{\dagger\mu_1}\ldots
  a^{\dagger\mu_{s-2}}\vac \,
  \overset{s}{\fr}_{\mu_1\ldots\mu_{s-2}},\notag
\end{align}
with totally symmetric tensor fields $\overset{s}{\varphi}$,
$\overset{s}{\fq}$, and~$\overset{s}{\fr}$.  The expression for
physical action~\eqref{eq:physaction}, with $\brst$ replaced by
$\brst_0$, then becomes
\begin{multline}
  \bundle{S}^{\mathrm{ph}}[\varphi,\fq,\fr] =-\half\int\! d^{\dmn}x\,
  \Big[ \inner{\Phi}{\cL\Phi} +\imath\inner{\Phi}{\cS^\dagger \FQ}
  -\imath\inner{\FQ}{\cS\Phi}\\*
  {}+\imath\inner{\FQ}{\cS^\dagger \FR} -\imath\inner{\FR}{\cS \FQ}
  +\inner{\FQ}{\FQ} -\inner{\FR}{\cL \FR} \Big],
\end{multline}
where $\inner{\cdot}{\cdot}$ is the inner product on the Fock space of
$a^\mu,a^{\dagger\mu}$ alone. Eliminating the auxiliary fields
contained in $\FQ$ and assuming reality conditions in accordance with
which the fields $\varphi_{\mu_1,\ldots,\mu_{s}}$ and
$\fr_{\mu_1,\ldots,\mu_{s-2}}$ are real, we arrive at
\begin{multline}\label{pre-Fro}
  \bundle{S}^{\mathrm{ph}}[\varphi,\fr] =-\int\! d^{\dmn}x\,
  \Big[\half\inner{\Phi}{\cL\Phi} - \half\inner{\cS\Phi}{\cS\Phi}+\\ +
  \inner{\FR}{\cS^2\Phi} - \inner{\FR}{\cL \FR} - \half\inner{\cS
    \FR}{\cS \FR} \Big].
\end{multline}
In component form, we rewrite this as
\begin{multline}
  \bundle{S}^{\mathrm{ph}}[\varphi,\fr] =-\sum_{s=0}^\infty
  \ffrac{1}{s!}  \int d^{\dmn}x\, \Big[ \half(\dd_\mu
  \overset{s}{\varphi},\dd^\mu\overset{s}{\varphi}) -
  \ffrac{s}{2}(\dd\cdot\overset{s}{\varphi},
  \dd\cdot\overset{s}{\varphi}) -
  \\
  {}- s(s-1)(\overset{s}{\fr},\dd\cdot\dd\cdot\overset{s}{\phi}) -
  s(s-1)(\dd_\mu\overset{s}{\fr},\dd^\mu\overset{s}{\fr})
  -\ffrac{s(s-1)(s-2)}{2} (\dd\cdot\overset{s}{\fr},
  \dd\cdot\overset{s}{\fr})\Big],
\end{multline}
where $(\dd\cdot\overset{s}{\varphi})_{\mu_1,\ldots,\mu_{s-1}}
=s\,\dd^\nu\overset{s}{\varphi}_{\nu\mu_1,\ldots,\mu_{s-1}}$ and the
standard inner product $(\cdot,\cdot)$ for symmetric tensors is
introduced, e.g., for $2$-tensors $a,b$ we have
$(a,b)=\eta^{\mu\rho}\eta^{\nu\sigma}a_{\mu\nu}b_{\rho\sigma}$.

The last expression is almost the sum of the Fronsdal Lagrangians
(which can of course be recognized directly in~\eqref{pre-Fro}).  To
obtain the Fronsdal action from~\eqref{pre-Fro}, it remains to impose
the conditions that the $\overset{s}{\fr}_{\mu_1,\ldots,\mu_{s-2}}$
fields are half the traces of
$\overset{s}{\varphi}_{\mu_1,\ldots,\mu_{s}}$ and that the
$\overset{s}{\varphi}_{\mu_1,\ldots,\mu_{s}}$ fields are
double-traceless. Remarkably, this can be done at the BRST
first-quantized level.  For this, we impose the additional constraint
\begin{equation}\label{eq:T-constr}
  T\equiv \eta_{\mu\nu}a^\mu a^\nu=0
\end{equation}
and let $\xi,\pi$, with
\begin{gather}
  \gh{\xi}=1,\qquad \gh{\pi}=-1,
\end{gather}
denote the corresponding ghost pair, represented in the space
$\cH_{\xi,\pi}$ of functions of~$\xi$.  The extended BRST operator is
then given by
\begin{equation} \label{eq:Fcharge}
  \brst=\brst_{-1}+\brst_0,\qquad
  \brst_{-1}=\xi \cT,\qquad \cT=\eta_{\mu\nu}a^\mu
  a^\nu+2cb,
\end{equation}
with $\brst_0$ defined in~\eqref{eq:Fbrst}.  With the degree taken to
be minus the homogeneity in $\xi$, the conditions of
Proposition~\bref{prop:red} are satisfied.  In fact, the cohomology
$H(\brst_{-1},\cH_0\tensor\cH_{\xi,\pi})$ can be identified with the
subspace $\cE$ of $\xi$-independent vectors $\psi$ satisfying
$\cT\psi=0$, because any symmetric tensor can be written as the trace
of a symmetric tensor.

It now follows from Proposition~\bref{prop:red} that after elimination
of the generalized auxiliary fields, the physical action for the BRST
system described by $\brst$ becomes
\begin{equation}
  \bundle{S}^{\mathrm{ph}}_F=-\half\int\, d^{\dmn}x\,
  \inner{\tilde\Psi^{(0)}}{\tilde\brst\tilde\Psi^{(0)}},
\end{equation}
where $\tilde \Psi$ is the string field associated with $\cE$ and
$\tilde\brst$ is the restriction of $\brst_0$ to $\cE$-valued
sections. Again, the fields associated with states with the ghost
dependence $\theta b^\dagger$ are ordinary auxiliary fields.  The
string field $\tilde \Psi^{(0)}(x)$ can be identified with the
solution of the constraint
\begin{equation}
  \cT\Psi^{(0)}=0
\end{equation}
for the string field of form~\eqref{eq:SF-phys}.  The solution is easy
to find in terms of the fields $\Phi$, $\FQ$, and~$\FR$,
\begin{equation}
  T\Phi=2\FR,\qquad  T \FQ=0, \qquad T \FR =0.
\end{equation}
Clearly, in terms of the tensor fields introduced in~\eqref{Phi-C-D},
$T$ is the operation of taking the trace.  After elimination of the
auxiliary fields $\overset{s}{\fq}_{\mu_1\ldots\mu_{s-1}}$, we thus
obtain the Fronsdal Lagrangians.

\subsubsection*{Remarks}\mbox{}

(i) We note that $\brst$ is no longer Hermitian in the standard inner
product on $\cH_0\tensor\cH_{\xi,\pi}$.  This is the reason why the
system described by $\brst$ is non-Lagrangian.  Nevertheless, as we
have just shown, it is equivalent to a Lagrangian system in the sense
of the previous subsection.

\medskip

(ii) In \cite{Bengtsson:1988jt} (see also \cite{Bengtsson:2004cd} for
a recent discussion), the Fronsdal Lagrangians were obtained by
imposing the constraint $\cT$ directly on the string field without
introducing a corresponding ghost pair and including it in a BRST
operator.  Other approaches to obtain the Fronsdal Lagrangians can be
found in~\cite{Pashnev:1998ti,Sagnotti:2003qa}.

\medskip

(iii) To describe the Fronsdal Lagrangian for a particular spin $s$,
the occupation-number constraint
\begin{equation}
  N_s\equiv a^\dagger_\mu a^\mu-c^\dagger b+b^\dagger c-s=0
\end{equation}
must be imposed in addition to $\cT=0$.  Trying to incorporate $N_s$
with some ghosts $\xi,\pi$ into the BRST operator $\brst_0$ in the
same way as we did for $\cT$, we see that each state annihilated by
$N_s$ appears twice in the cohomology, once multiplied by the ghost
$\xi$ and once without it.  This is why the individual Fronsdal
Lagrangians cannot be directly obtained by eliminating generalized
auxiliary fields, not even on the level of the equations of motion.  A
way out is to treat the constraint $N_s=0$ in the same way as the
level-matching condition in closed string field theory
\cite{Zwiebach:1993ie}: the doubling is eliminated by requiring the
string field to be annihilated by $\pi$, and the action is obtained by
regularizing the inner product with an insertion of $\xi$.  An
alternative is not to introduce ghost pairs for $\cT$ and $N_s$ in the
first place and impose both $\cT=0$ and $N_s=0$ as constraints on the
string field.  This can be done consistently because
$\commut{N_s}{\cT}=-2\cT$ and both constraints commute~with~$\brst_0$.

\section{The first-order parent system}
\label{sec:fedosov}
We now show how to replace a BRST first-quantized system by an
equivalent extended system that is first-order in space--time
derivatives.  Such a construction is related to the conversion of
second-class constraints~\cite{Batalin:1987fm,Batalin:1990mb} or a
version of Fedosov quantization~\cite{Fedosov:1996fu} (see
\cite{Grigoriev:2000rn,Batalin:2001je} for a unified description of
both).  The field theory associated with the extended system is then
also first-order and is reduced to the original one by elimination of
generalized auxiliary fields.  A different reduction of the extended
theory by the elimination of another collection of generalized
auxiliary fields gives rise to a first-order system representing the
unfolded form~\cite{Vasiliev:1988xc,Vasiliev:1994gr} of the original
field theory.  This last reduction is explicitly illustrated in the
case of the Klein--Gordon and Fronsdal equations.

\subsection{Extended first-quantized system and parent field theory}
We recall the classical constrained system on $T^*\manX\times B$
described at the beginning of Sec.~\bref{subsec:1st-quant}, where
$x^{\mu},p_{\mu}$, ${\mu}=1,\dots,\dmn$, are coordinates
on~$T^*\manX=T^*\oR^\dmn$. \ We embed the phase space $T^*\manX\times
B$ into the extended phase space $T^*(T\manX)\times B$ as the
second-class constraint surface determined~by
\begin{equation}\label{constr-add}
  p_{\mu}-p^y_{\mu}=0,
  \qquad y^{\mu}=0.
\end{equation}
In terms of coordinates $x^{\mu},y^{\nu},{p_{\mu}},{p^y_{\nu}}$ on
$T^*(T\manX)$, the corresponding canonical Poisson brackets are
\begin{equation}
  \pb{ p_{\mu}}{ x^{\nu}}=-\delta^{\nu}_{\mu},
  \qquad \pb{p^y_{\mu}}{y^{\nu}}=-\delta_{\mu}^{\nu}.
\end{equation}
Constraints~\eqref{constr-add} are then indeed second-class.  The
Poisson bracket on $T^*(T\manX)\times B$ is given by the product of
the Poisson brackets of the factors.  In the BRST formalism, these
constraints can be taken into account by treating only the first set
as first-class constraints.  This requires extending the phase space
further by ghosts $\cC^{\mu}$ and ghost momenta $\cP_{\mu}$ with the
Poisson bracket $\pb{\cP_{\mu}}{ \cC^{\nu}}=-\delta^{\nu}_{\mu}$.

At the quantum level, the phase-space coordinates become operators
with the nonzero commutation relations
\begin{equation}\label{commutators-after}
  \commut{p_{\nu}}{x^{\mu}}
  =-\imath\delta^{\mu}_{\nu},\qquad
  \commut{p^y_{\nu}}{y^{\mu}}=-\imath\delta^{\mu}_{\nu},\qquad
  \commut{\cP_{\nu}}{\cC^{\mu}}=-\imath\delta^{\mu}_{\nu}.
\end{equation}
As in Sec.~\bref{subsec:1st-quant}, we here assume that the internal
space $B$ is quantized, with the representation space denoted
by~$\cH$.  The BRST operator corresponding to
constraints~\eqref{constr-add}~is
\begin{equation}
  \label{eq:brst-T0}
  \brst^{\manX}=\cC^{\mu}( p_{\mu}- p^y_{\mu}).
\end{equation}

All the constraints, including the original ones on~$T^*\manX\times B$
(Eqs.~\eqref{constr-ori}), are taken into account by the ``total''
BRST operator
\begin{equation}\label{brstt-first}
  \brst^{\T}=\brst^{\manX}+\brst^y,
\end{equation}
where $\brst^y$ is constructed as a formal power series in $y^\mu$
and~$p^y_{\mu}$ such that
\begin{equation}
  \commut{\brst^{\manX}+\brst^y}{\,\brst^{\manX}+\brst^y}=0,
  \qquad \brst^y|_{y=0,\, p_{\mu}-p^y_{\mu}=0}=\brst.
\end{equation}
The construction of $\brst^{\T}$ is an adapted version of the general
method of including original first-class constraints in an extended
system, see~\cite{Batalin:1990mb}.

In what follows, \textit{we assume BRST operator $\brst$ to be
  $x$-independent}, and hence $\brst^y$ can be taken to be just
$\brst$ with $p_\mu$ replaced by $p^y_\mu$.

The space of quantum states of the extended system is chosen to be the
space $\Gamma(\bundle{\cH}^{\T})$ of $\cH^{\T}$-valued sections, with
$\cH^{\T}=\cH\tensor\Salgebra(T^*)\tensor\wwedge{(T^*)}$, where $T^*$
is the cotangent space to $\manX$ and where $\Salgebra$ and $\wwedge$
denote the symmetric and skew-symmetric tensor algebras.  In other
words, states can be identified with ``functions'' of the form
$\bundle{\phi}=\bundle{\phi}(x,y,\cC)=e_A\bundle{\phi}^A(x,y,\cC)$,
where $(e_A)$ is a basis in $\cH$ and each $\bundle{\phi}^A$ is a
function in $x$, a formal power series in the commuting variables
$y^{\mu}$, and a polynomial in the anticommuting
variables~$\cC^{\mu}$, ${\mu}=1,\dots,\dmn$.  The momenta $p_{\mu}$,
$p^y_{\mu}$, and the ghost momenta $\cP_{\mu}$ act as
$-\imath\dl{x^{\mu}}$, $-\imath\dl{y^{\mu}}$, and
$-\imath\dl{\cC^{\mu}}$ respectively; as noted above, the operators
corresponding to the internal degrees of freedom are represented on
$\cH$.  The basis elements in $\cH^{\T}$ are given~by
\begin{equation}
  \label{eq:T-basis}
  e^{({\mu})[{\nu}]}_{A}
  =e_A \tensor y^{{\mu}_1}\ldots y^{{\mu}_n}\,
  \cC^{{\nu}_1}\ldots \cC^{{\nu}_l},
\end{equation}
where $({\mu})$ and $[{\nu}]$ denote a collection of respectively
symmetrized and antisymmetrized indices.

Following Sec.~\bref{subsec:1st-quant}, we construct the field theory
for the extended system $(\brst^{\T},\Gamma(\bundle{\cH}^{\T}))$ in
terms of the string field
\begin{equation}
  \Psi^{\T}(x)=e_A^{({\mu})[\nu]}\psi^A_{({\mu})[\nu]}(x).
\end{equation}
The total BRST operator $\brst^{\T}$ acts on $\Psi^{\T}$ as
\begin{equation} \label{brstt}
  \brst^{\T}\Psi^{\T}(x)=-\imath (\derham
  +\sigma+\imath\brst^y)\Psi^{\T}(x),
\end{equation}
where
\begin{equation}\label{rhosigma}
  \derham=\cC^\mu\dl{x^\mu},\qquad
  \sigma=-\cC^{\mu}\dl{y^{\mu}}
\end{equation}
(and we recall that $\brst^y$ is given by $\brst$ with the momenta
$p_\mu$ replaced by $p^y_\mu$, which act as $-\imath\dl{y^{\mu}}$).
The BRST differential $s^{\T}$ is then determined by
$s^{\T}\Psi^{\T}=\brst^{\T}\Psi^{\T}$.

In what follows, the field theory determined by $s^{\T}$ and
$\Psi^{\T}$ is called the \textit{parent theory}.  Its role is to
produce apparently different theories via elimination of different
collections of generalized auxiliary fields.  By the above analysis,
all these theories are guaranteed to be just equivalent
representations of the same dynamics.  Two of its reductions are
studied in Secs.~\bref{sec:red-ori} and~\bref{sec:red-unfolded}.

\subsection{Reduction to the original system}\label{sec:red-ori}
We now show that via the elimination of generalized auxiliary fields,
the parent theory determined by the BRST differential $s^T$ is
equivalent to the original theory determined by the BRST differential
$s\Psi^A = (\brst\Psi)^A$.

\begin{prop}\label{prop:parent2ori}
  The parent system $(\brst^{\T},\Gamma(\bundle{\cH}^{\T}))$ can be
  algebraically reduced to the original, $y^\mu$- and
  $\cC^\mu$-independent system $(\brst,\Gamma(\bundle\cH))$.
\end{prop}
\begin{proof}
  The proof consists in (i) constructing a decomposition
  \begin{equation}\label{H-EFG}
    \cH^{\T}=\cE\oplus\cF\oplus\cG
  \end{equation}
  where $\cE$ is isomorphic to $\cH$ with $\st{\cG\cF}{\brst^{\T}}$
  invertible, and (ii) showing that $\widetilde{\brst^{\T}}$ is mapped
  to $\brst$ under the isomorphism.
  
  To construct the required decomposition of $\cH^{\T}$, we use
  Proposition~\bref{prop:red} with the underlying grading chosen as
  follows.  Our assumptions imply that $\brst^y$ is a polynomial in
  the $\dl{y^\mu}$, $\mu=1,\dots,\dmn$.  Let $\ell$ be the maximum
  power of $\dl{y^\mu}$ involved in $\brst^y$.  The grading is given
  by $\text{(homogeneity degree in $y$)}+\ell\,(\text{target-space
    ghost number})$, where the \textit{target-space ghost number} is
  the ghost number that does not count the number of $\cC^\mu$'s. This
  grading is assumed to be bounded from below and, without loss of
  generality, the bound can be taken to be zero.
  
  With respect to this grading (indicated by a subscript), the BRST
  operator $\brst^{\T}$ decomposes as in~\eqref{eq:2diffeq} with
  $\imath\brst^{\T}_{-1}=\sigma$.  It is then obvious that the
  cohomology of $\brst^{\T}_{-1}$ in $\cH^{\T}$ can be identified with
  $\cH\subset \cH^{\T}$.
  
  To complete decomposition~\eqref{H-EFG}, we introduce the operators
  \begin{equation}
    \rho=-y^\mu\dl{\cC^\mu},\qquad \tau=y^\mu\dl{x^\mu},
    \qquad
    N=\cC^\mu\dr{\cC^\mu}+y^\mu\dr{y^\mu}
  \end{equation}
  acting in~$\cH^{\T}$ and set
  \begin{equation}
    \cF=\rho\,
    \cH^{\T},
    \qquad
    \cG=\sigma\,\cH^{\T}.
  \end{equation}
  By Proposition~\bref{prop:red}, the operator
  $\st{\cG\cE}{\brst^{\T}}$ is algebraically invertible, and therefore
  the parent system $(\brst^{\T},\Gamma(\bundle{\cH}^{\T}))$ can be
  algebraically reduced to the system
  $(\widetilde{\brst^{\T}},\Gamma(\bundle{\cH}))$.
  
  It remains to calculate $\widetilde{\brst^{\T}}$.  For this, we
  introduce the grading with respect to the space-time ghost number,
  which is determined by the operator $\cC^\mu\dl{\cC^\mu}$.  With
  this grading indicated by a superscript, we have the decomposition
  \begin{equation}
    \brst^{\T}= \brst_{-1}^1+\brst_{0}^1
    +{}\smash[b]{\sum_{i=0}^\ell\brst_i^0},
  \end{equation}  
  where it is readily seen that
  \begin{gather}
    \imath \brst_{-1}^1=\sigma,\qquad \imath \brst_{0}^1=\derham.
  \end{gather}
  Moreover, $\brst_i^0=\brst^y_{[\ell-i]}$, where $\brst^y_{[m]}$ is
  the term in $\brst^y$ that is homogeneous of degree $m$ in
  $\dl{y^\mu}$ and we set $\brst^0=\sum_{i=0}^\ell\brst_i^0$.
  
  The space $\cH^{\T}$ decomposes accordingly,
  \begin{gather}
    \cH^{\T}=\smash[b]{\bigoplus_{i\geq0}\bigoplus_{j=0}^{\dmn}}
    (\cH^{\T})_{i}^{j}, \intertext{and it follows that}
    \cE=\bigoplus_{i\geq 0}\cE_{i\ell}^0\,.
    \label{E-0il}
  \end{gather}
  We refer to (elements of) $(\cH^{\T})_{i}^{j}$ as having the
  bidegree~$\binom{j}{i}$.
  
  To construct $\widetilde{\brst^{\T}}$, we note that
  \begin{gather}
    \st{\cE\cE}{\brst^{\T}}=\brst_\ell^0=\brst^y_{[0]},\qquad
    \imath\st{\cG\cE}{\brst^{\T}}=\imath\brst_0^1=\derham.
  \end{gather}
  We also note that $N$ is invertible on the image of~$\rho$ and on
  the image of~$\sigma$.  According to~\eqref{eq:2}, the operator
  $(\st{\cG\cF}{\brst^{\T}})^{-1}:\cG\to\cF$ is given by
  \begin{multline}
    (\imath\st{\cG\cF}{\brst^{\T}})^{-1}\bundle{\phi}^\cG =[N^{-1}\rho
    -N^{-1}\rho(\derham+\imath\brst^0)
    N^{-1}\rho\\+N^{-1}\rho(\derham+\imath\brst^0) N^{-1}
    \rho(\derham+\imath\brst^0)N^{-1}\rho-\dots]
    \bundle{\phi}^\cG,\qquad \forall \bundle{\phi}^\cG\in
    \Gamma(\bundle{\cG}).
  \end{multline}
  and is a two-sided inverse of~$\imath\st{\cG\cF}{\brst^{\T}}$.  A
  simple argument based on grading shows that on any
  $\bundle{\phi}^\cE$ in $\Gamma(\bundle{\cE})$,
  \begin{multline}
    (\imath\st{\cG\cF}{\brst^{\T}})^{-1} \imath\st{\cG\cE}{\brst^{\T}}
    \bundle{\phi}^\cE = \Bigl(N^{-1}\rho -N^{-1}\rho\derham
    N^{-1}\rho+N^{-1}\rho\derham N^{-1}\rho\derham N^{-1}\rho-\dots
    \Bigr)\derham\, \bundle{\phi}^\cE\\*
    =\sum_{n\geq1}(-1)^{n-1}\bigl(N^{-1}\rho\derham\bigr)^n
    \bundle{\phi}^\cE.
  \end{multline}
  The term $\bigl(N^{-1}\rho\derham\bigr)^n$ increases the bidegree by
  $\binom{0}{n}$.  Because $\cE$ has bidegree~$\binom{0}{i\ell}$
  (see \eqref{E-0il}), it follows that
  \begin{equation}
    \imath\st{\cE\cF}{\brst^{\T}}
    (\imath\st{\cG\cF}{\brst^{\T}})^{-1} \imath\st{\cG\cE}{\brst^{\T}}
    \bundle{\phi}^\cE 
    = 
    \sum_{n\geq1}(-1)^{n-1}
    (\imath\st{\cE\cF}{\brst^{\T}})^{0}_{\ell-n}
    \bigl(N^{-1}\rho\derham\bigr)^n \bundle{\phi}^\cE.
  \end{equation}
  Noting that $\smash{(\st{\cE\cF}{\brst^{\T}})^{0}_{\ell-n}}
  =\brst^0_{\ell-n}=\brst^y_{[n]}$, we finally
  evaluate~\eqref{tilde-brst} as
  \begin{equation}
    \tilde{\brst^{\T}}\bundle{\phi}^\cE=\brst^y_{[0]}
    +\sum_{n=1}^{\ell}(-1)^{n}\,\brst^y_{[n]} 
    (N^{-1}\rho\derham)^{n}\bundle{\phi}^\cE.
  \end{equation}
  Rearranging the operators in this formula using that $\rho\cE=0$ and
  $\commut{\rho}{\derham}=-\tau$, we obtain
  \begin{equation}\label{brst-decompos}
    \tilde\brst^{\T}\bundle{\phi}^\cE
    =\sum_{n=0}^\ell\ffrac{1}{n!}\brst^y_{[n]}\tau^{n}\bundle{\phi}^\cE.
  \end{equation}
  For $n\geq1$, moreover, $\brst^y_{[n]}\bundle{\phi}^\cE=0$ and
  $\brst^y_{[n]}\tau^{n}\bundle{\phi}^\cE=
  n!\brst_{[n]}\bundle{\phi}^\cE$, where $\brst_{[n]}$ is the term in
  $\brst$ that is homogeneous of degree $n$ in $\dl{x^\mu}$.
  Therefore, $\tilde\brst^{\T}\bundle{\phi}^\cE=\brst
  \bundle{\phi}^\cE$.
\end{proof}

\subsection{Reduction to the unfolded formalism}
\label{sec:red-unfolded}
Taking another reduction of the parent system
$(\brst^{\T},\Gamma(\bundle{\cH}^{\T}))$ to the subspace $\cE\simeq
H(\brst^y,\cH^{\T})$, we now construct the unfolded form of the
original system.

\begin{prop}\label{prop:parent2unfold}
  The parent system $(\brst^{\T},\Gamma(\bundle{\cH}^{\T}))$ can be
  algebraically reduced to
  $(\tilde\brst^{\T},$\linebreak[]0$\Gamma(\bundle\cE))$ with
  $\cE\simeq H(\brst^y,\cH^{\T})$.
\end{prop}

\begin{proof}
  To prove this, we take minus the target-space ghost number (defined
  in the paragraph after Eq.~\eqref{H-EFG}) as the degree.  The BRST
  operator $\brst^{\T}$ in~\eqref{brstt-first} decomposes as
  $\brst^{\T}_{-1}=\brst^y$, $\brst^{\T}_0=\brst^{\manX}$.  We assume
  the target-space ghost number to be bounded from above and hence the
  degree to be bounded from below.  Without loss of generality, this
  bound can again be assumed to be zero.  The graded
  finite-dimensional space $\cH^{\T}$ can be decomposed as
  $\cH^{\T}=\cE\oplus\cF\oplus\cG$ with $\Ker \brst^y\supset\cE\simeq
  H(\brst^y,\cH^{\T})$ and $\cG=\im(\brst^y)$.  Because the conditions
  of Proposition~\bref{prop:red} are then satisfied,
  $\st{\cG\cF}{\brst^{\T}}$ is invertible, and the system can be
  reduced to $(\tilde\brst^{\T},\Gamma(\bundle{\cE}))$.
\end{proof}

Explicitly, it follows from~\eqref{tilde-brst} that the reduced BRST
operator $\tilde\brst^{\T}$ is given by
\begin{equation}\label{it-follows}
  \imath\tilde\brst^{\T}=\derham+\widetilde\sigma,\qquad
  \widetilde\sigma=\st{\cE\cE}{\sigma}
  -\st{\cE\cF}{\sigma}(\st{\cG\cF}{\sigma}
  +\imath\st{\cG\cF}{\brst^y})^{-1}
  \st{\cG\cE}{\sigma}.
\end{equation}
Here, we have used that $\derham\Gamma(\bundle{\cE})\subset
\Gamma(\bundle{\cE})$, and we also have used $\derham$ to denote the
restriction of $\derham$ to $\Gamma(\bundle{\cE})$.  Similar
conventions are used for operators that restrict to a subspace.

The inverse of $\st{\cG\cF}{\sigma}+\imath\st{\cG\cF}{\brst^y}$
involved in~\eqref{it-follows} is constructed as follows.  If
$\rho\map \cG\to\cF$ is the inverse to $\st{\cG\cF}{\imath\brst^y}\map
\cF \to \cG$, then
\begin{equation}
  (\st{\cG\cF}{\sigma}+\imath\st{\cG\cF}{\brst^y})^{-1} =
  \rho-\rho\st{\cG\cF}{\sigma}\rho
  +\rho\st{\cG\cF}{\sigma}\rho\st{\cG\cF}{\sigma}\rho-\ldots.
\end{equation}
Accordingly, $\tilde\sigma\map\cE\to\cE$ takes the form
\begin{equation}\label{accordingly}
  \tilde\sigma=
  \st{\cE\cE}{\sigma} - \st{\cE\cF}{\sigma}\rho\st{\cG\cE}{\sigma} +
  \st{\cE\cF}{\sigma}\rho\st{\cG\cF}{\sigma}\rho\st{\cG\cE}{\sigma}
  -
  \st{\cE\cF}{\sigma}\rho\st{\cG\cF}{\sigma}
  \rho\st{\cG\cF}{\sigma}\rho\st{\cG\cE}{\sigma}
  +\ldots.
\end{equation}
The $\ell$th term in this expansion has target-space ghost number
$1-\ell$ and form degree $\ell$.  In particular, all terms after the
$d$th term vanish because there cannot be terms of form degree higher
than $d$, the dimension of $\manX$.  Furthermore, if the
$\brst^y$-cohomology vanishes below target-space ghost number
$k_{\min}$ and above target-space ghost number $k_{\max}$, then all
terms after the $(k_{\max}-k_{\min}+1)$th term also vanish.

With $\tilde\Psi^{\T}$ denoting the string field associated with
$\cE$, the equations of motion for the reduced system are given by
\begin{equation} \label{eq:UEOM}
  (\derham+\widetilde\sigma)\tilde\Psi^{\T (0)}=0,
\end{equation}
with gauge transformations described by
\begin{equation}
  \delta \tilde\Psi^{\T (0)}
  =(\derham+\widetilde\sigma)\tilde\Psi^{\T (1)},\label{4.18}
\end{equation}
where the ghost-number-$1$ component fields of $\tilde\Psi^{\T (1)}$
are to be interpreted as gauge parameters.  We claim that
equations~\eqref{eq:UEOM} are the unfolded form of the original
equations in the sense
of~\cite{Vasiliev:1994gr,Vasiliev:1999ba,Shaynkman:2004vu}.

\subsection{From the parent theory to the unfolded formulation.
  Examples}
We illustrate the general method of reducing the parent system to its
unfolded form as in Sec.~\bref{sec:red-unfolded} with two examples,
the relativistic particle and the Fronsdal system of higher-spin
fields.

\subsubsection*{Klein--Gordon equations}\label{sec:KG}
We first consider the simplest example, with the parent system the
relativistic particle described by the BRST operator $\brst=\theta
\eta^{\mu\nu}p_\mu p_\nu$, whence $\brst^y=\theta\eta^{\mu\nu}p^y_\mu
p^y_\nu$.  The space $\cH^{\T}$ can be viewed as the space of formal
power series in $y^\mu$ and $\theta$, $\cC^\mu$.  The cohomology of
$\brst^y$ in $\cH^{\T}$ can easily be shown to be represented by the
space $\cE\subset\cH^{\T}$ of $\theta$-independent formal power series
in $y^\mu,\cC^\mu$ corresponding to traceless tensors in $y^\mu$,
i.e., the series satisfying the constraint
\begin{equation}
  \eta_{\mu\nu}\dl{y^\mu}\dl{y^\nu}f(y,\cC)=0.
\end{equation}
With the degree being minus the target-space ghost number (i.e., minus
the homogeneity degree in~$\theta$ in our case), the cohomology is
concentrated in one degree and we see that
$\tilde\sigma=\st{\cE\cE}{\sigma}=\cC^\mu\tilde p^y_\mu$, where
$\tilde p^y_\mu$ is the action of $p^y_\mu$ in $\cE$.  Next, because
there are no states in negative target-space ghost numbers, the
physical fields, which are component fields of $\tilde\Psi^{\T (0)}$,
are associated with $\cC^\mu$-independent states and are therefore
traceless symmetric tensor fields.  The equations of motion then take
the form
\begin{equation}
  (\dd_\mu - \tilde p^y_\mu)\tilde\Psi^{\T (0)}=0.
\end{equation}
These equations are the unfolded form of the Klein--Gordon equation
(see~\cite{Shaynkman:2000ts}).

We note that in all models where the cohomology of $\brst^y$ vanishes
in negative ghost numbers, as is the case in this example, it follows
from~\eqref{4.18} that the component fields contained in
$\tilde\Psi^{\T (0)}$ are gauge invariant, because there are no gauge
transformations that affect them.

\subsubsection*{Fronsdal equations} \label{sec:U-Fronsdal}
As another example of the reduction in Sec.~\bref{sec:red-unfolded},
we now construct the unfolded form of the Fronsdal system described in
Sec.~\bref{sec:Fronsdal}.

For the BRST operator given by~\eqref{eq:Fcharge}
and~\eqref{eq:Fbrst}, the operator $\brst^y$ is obtained by replacing
$p_\mu$ with $p^y_\mu$ (see~\eqref{commutators-after}).  The
representation space $\cH^{\T}$ can be taken as the space of formal
power series in the variables $y^\mu$ and polynomials in
$a^{\dagger\mu}$ and in the ghosts
$\cC^{\mu},\eta,\xi,c^\dagger,b^\dagger$.  The operator
$\imath\brst^y$ acts in the representation space as
\begin{equation} \label{eq:omega-y}
  \imath\brst^y
  =-\imath\theta\Box+c^\dagger\yS+\yS^\dagger \dl{b^\dagger}
  -\imath c^\dagger \dl{b^\dagger}\dl{\theta}
  +\imath\xi T-2\imath\xi \dl{b^\dagger}\dl{c^\dagger},
\end{equation}
where we introduce the notation
\begin{equation}\label{yS-here}
  \Box=\eta^{\mu\nu}\dl{y^\mu}\dl{y^\nu},\quad
  \yS=\eta^{\mu\nu}\dl{a^{\dagger\mu}}\dl{y^\nu}, \quad
  \yS^\dagger=a^{\dagger\mu}\dl{y^\mu},\quad
  T=\eta^{\mu\nu}\dl{a^{\dagger\mu}}\dl{a^{\dagger\nu}}.
\end{equation}
We remind the reader that the total BRST operator is
\begin{gather}
  \imath \brst^{\T}=\derham+\sigma+\imath{\brst^y}
\end{gather}
with $\derham$ and $\sigma$ defined in~\eqref{rhosigma}.

To calculate the result of the reduction described in
Sec.~\bref{sec:red-unfolded}, it is convenient to proceed in two
steps.  At the first step, we obtain an \textit{intermediate system},
whose further reduction gives the unfolded form of the Fronsdal
equations.

The total BRST operator $\brst^{\T}$ splits as in~\eqref{eq:2diffeq},
with the underlying grading given by minus the homogeneity degree in
$\theta$, $c^\dagger$, and $\xi$:
\begin{equation}\label{T-split}
  \begin{gathered}
    \brst^{\T}=\brst^{\T}_{-1}+\brst^{\T}_{0},
    \\
    \imath\brst^{\T}_{-1} =-\imath\theta\Box +c^\dagger\yS +\imath\xi
    T, \quad \brst^{\T}_{0} =\derham +\sigma +\yS^\dagger
    \dl{b^\dagger} -\imath c^\dagger \dl{b^\dagger}\dl{\theta}
    -2\imath\xi \dl{b^\dagger}\dl{c^\dagger}.
  \end{gathered}
\end{equation}
The first step of the reduction is to the cohomology of
$\brst^{\T}_{-1}$, which we now describe.
\begin{lemma}\label{lemma:1st-step}
  \mbox{}
  \begin{enumerate}
  \item For $\dmn=1,2$, \ $H_0(\brst^{\T}_{-1},\cH^{\T})\neq0$,
    $H_1(\brst^{\T}_{-1},\cH^{\T})\neq0$, and
    $H_i(\brst^{\T}_{-1},\cH^{\T})=0$ for $i\geq2$.
    
  \item For $\dmn\geq3$, $H_0(\brst^{\T}_{-1},\cH^{\T})\neq0$ and
    $H_i(\brst^{\T}_{-1},\cH^{\T})=0$ for $i\geq1$.
  \end{enumerate}
  Moreover, in both cases, the space $H_0(\brst^{\T}_{-1},\cH^{\T})$
  is isomorphic to the space $\hat \cE$ of elements
  $\Phi(y,a^{\dagger},b^\dagger,\cC)$ of $\cH^\T$ satisfying the
  relations
  \begin{equation} \label{eq:hatcE-cond}
    \Box \Phi=0,\qquad \yS \Phi=0, \qquad T\Phi=0.
  \end{equation}    
\end{lemma}
\begin{proof}
  The cohomology is to be calculated in the space of functions that
  are formal power series in the variables $y^\mu$ and polynomials in
  $a^{\dagger\mu}$, $\cC^\mu$, $b^\dagger$, $\theta$, $c^\dagger$, and
  $\xi$.  The variables $\cC^\mu$ and $b^\dagger$ and their conjugates
  are not involved in $\brst^{\T}_{-1}$, and hence they enter the
  cohomology as tensor factors.  We omit this dependence in the rest
  of the proof.  Next, $\imath\brst^{\T}_{-1}$ is homogeneous in
  $a^{\dagger\mu}$ and $y^\mu$, of degree $-2$.  This implies that the
  cohomology can be calculated on the polynomials that have a definite
  total homogeneity degree in $(y^\mu, a^{\dagger\mu})$.  We can thus
  evaluate the cohomology in the space of polynomials in $y^\mu$,
  $a^{\dagger\mu}$, $\theta$, $c^\dagger$, and~$\xi$.
  
  The BRST operator consists of three pairwise commuting parts, and
  hence its cohomology can be evaluated as the cohomology of
  $c^\dagger\yS$ on the cohomology of $\xi T$ evaluated on the
  cohomology of~$\theta\Box$.  The cohomology of $\xi T$ on the
  cohomology of~$\theta\Box$ is given by the space of all polynomials
  in $y^\mu$, $a^{\dagger\mu}$, and $c^\dagger$ that are annihilated
  by~$T$ and~$\Box$. Indeed, because all polynomials in~$y^\mu$ can be
  written as~$\Box$ acting on some polynomial in~$y^\mu$, the
  cohomology of~$\theta\Box$ reduces to $\theta$-independent
  polynomials in~$y^\mu$ that are in the kernel of~$\Box$.  The same
  reasoning can be applied for the cohomology of~$\xi T$ in the
  cohomology of~$\theta\Box$.  We are therefore left with the
  ``utmost'' cohomology of~$c^\dagger\yS$.  Let $\spaceN=\ker
  T\cap\ker\Box$ on the space of polynomials in~$y^\mu$
  and~$a^{\dagger\mu}$ and let~$\yS_\spaceN$ denote the restriction
  of~$\yS$ to~$\spaceN$.
  
  We note that whenever $\yS_\spaceN$ has the property that
  $\im\yS_\spaceN=\spaceN$, the cohomology is concentrated in degree
  zero and is given by $\ker\yS_\spaceN\subset\spaceN$, and is
  therefore described by conditions~\eqref{eq:hatcE-cond}.  But this
  property of $\yS_\spaceN$ is proved in Appendix~\bref{sec:A} for
  $\dmn\geq3$, which shows assertion~(2).  The proof in the Appendix
  is based on the relation of the BRST operator to the $sp(4)$ algebra
  and involves standard representation-theory techniques that allow
  finding the occurrences of singular vectors in Verma and generalized
  Verma modules.  In short, the proof amounts to showing that for
  $\dmn\geq3$, \ $sp(4)$ Verma modules contain no singular vectors
  except the one generated by a suitable power of the operator
  $\bar\yS^\dagger$.  For $d=1,2$, the analysis shows that
  $\yS_\spaceN$ has a nonzero cokernel in~$\spaceN$ because of the
  existence of additional singular vectors; finding their number and
  positions then yields assertion~(1).
\end{proof}
We restrict to the case $\dmn\geq3$ in what follows.

By proposition~\bref{prop:red}, the system $(\brst^{\T},
\Gamma(\bundle{\cH}^{\T}))$ can be algebraically reduced to
$(\hat\brst^\T,$\linebreak[0]$\Gamma(\bundle{\hat\cE}))$ with
\begin{gather}\label{brstt-red1}
  \imath \hat\brst^{\T} = \derham +\sigma +\yS^\dagger \dl{b^\dagger}.
\end{gather}

Let ${\hat\Psi}^{\T}$ be the string field associated with the subspace
$\hat\cE$.  Its ghost-number-zero component is
\begin{equation}
  {\hat\Psi}^{\T (0)}=\hat\Psi^0+ b^\dagger\hat\Psi^1,
  \qquad
  \hat\Psi^1=\cC^\mu\hat\Psi^1_\mu
\end{equation}
(where $\hat\Psi^0$ and $\hat\Psi_\mu^1 $ are associated with the
subspaces in $\hat\cE$ of the respective target-space ghost numbers
$0$ and~$-1$).  Hence, we have the following proposition.
\begin{prop}
  The parent field theory for the Fronsdal free higher-spin gauge
  theory can be reduced to the field theory associated with the
  intermediate system $(\hat\brst^{\T},\Gamma(\bundle{\hat\cE}))$ by
  elimination of generalized auxiliary fields. The associated
  equations of motion are explicitly given by
  \begin{equation}
    \begin{split}
      (\derham + \sigma)\hat\Psi^0&=-\yS^\dagger \hat\Psi^1,\\
      (\derham + \sigma)\hat\Psi^1&=0.
    \end{split}
  \end{equation}
\end{prop}
The main feature of the intermediate system
$(\hat\brst^T,\Gamma(\bundle{\hat\cE}))$ is that the states (and the
associated fields) are only constrained by trace
conditions~\eqref{eq:hatcE-cond} and that the reduced BRST operator
remains simple after the elimination of a considerable amount of
states (which correspond to algebraically contractible pairs).

Having obtained the intermediate system
$(\hat\brst^{\T},\Gamma(\bundle{\hat\cE}))$, we proceed with the
second step of the reduction. Following the strategy described in
Sec.~\bref{sec:red-unfolded}, we now perform the reduction from
$(\hat\brst^{\T},\Gamma(\bundle{\hat\cE}))$ to another system
$(\tilde\brst^{\T},\Gamma(\bundle{\cE}))$ using
decomposition~\eqref{eq:2diffeq} again, this time with the underlying
grading given by minus the target-space ghost number, which is in the
case of $\hat\cE$ the homogeneity degree in~$b^\dagger$:
\begin{gather}\label{decomp-next}
  \hat\brst^{\T}_{-1} = \yS^\dagger \dl{b^\dagger}, \qquad
  \hat\brst^{\T}_{0} =\derham +\sigma.
\end{gather}
To describe the reduced system, we must determine the cohomology of
$\hat\brst^{\T}_{-1}$ and evaluate the reduction of the BRST operator.

We first find the cohomology of $\brst^{\T}_{-1}$.  We recall that
$\dmn\geq3$ and also recall the notation
$\bar\yS^\dagger=y^\mu\dl{a^{\dagger\mu}}$ from~\eqref{all-sp4} for a
generator of the $s\ell(2)$ algebra in~\eqref{the-sl2}.
\begin{lemma}\label{lemma:2nd-step} In degree $0$,
  $H_0(\hat\brst^{\T}_{-1},\hat\cE)$ is isomorphic to the space
  $\cE_0$ of elements from $\cH^\T$ of the form
  $\Phi(y,a^{\dagger},\cC)$ satisfying the relations
  \begin{gather}\label{E0-def}
    \Box \phi=0,\quad \yS \phi=0, \quad T\phi=0, \quad
    \bar\yS^\dagger\phi = 0
  \end{gather}
  and $H_1(\hat\brst^{\T}_{-1},\hat\cE)$ is isomorphic to the space
  $\cE_1$ of elements in $\cH^\T$ of the form
  $b^\dagger\lambda(y,a^{\dagger},\cC)$ satisfying the relations
  \begin{gather} \label{eq:Lambda-cocycle}
    \Box\lambda=0,\quad \yS\lambda=0, \quad
    T\lambda=0,\quad\yS^\dagger\lambda=0.
  \end{gather}
\end{lemma}
In what follows, we use the notation $\cE=\cE_0\oplus\cE_1\simeq
H(\hat\brst^\T_{-1},\hat\cE)$.

We note that the elements of $\hat \cE$ specified in the proposition
depend on $\cC^\mu$ arbitrarily; as functions of $y^\mu$ and
$a^{\dagger\mu}$, they decompose into monomials. Those singled out
by~\eqref{E0-def} correspond to traceless tensors described by the
Young tableaux
\begin{equation} \label{eq:YT1}
  \begin{picture}(100,25)
    \put(0,-10){
      \put(-12,25){\tiny $y$}
      \put(-12,10){\tiny $a^\dagger$}
      \put(0,30){\line(1,0){90}}
      \put(0,20){\line(1,0){90}}
      \put(90,20){\line(0,1){10}}
      \put(80,20){\line(0,1){10}}
      \put(63,21.5){$\cdots$}
      \put(60,20){\line(0,1){10}}
      \put(50,10){\line(0,1){20}}
      \put(0,10){\line(1,0){50}}
      \put(40,10){\line(0,1){20}}
      \put(16.5,21.5){$\cdots$}
      \put(16.5,11.5){$\cdots$}
      \put(10,10){\line(0,1){20}}
      \put(0,10){\line(0,1){20}}
    }
  \end{picture}
\end{equation}
(where the length of the first row is greater than or equal to the
length of the second); the monomials singled out
by~\eqref{eq:Lambda-cocycle} correspond to traceless tensors described
by the Young tableaux
\begin{equation} \label{eq:YT2}
  \begin{picture}(100,25)
    \put(0,-10){
      \put(-12,23){\tiny $a^\dagger$}
      \put(-12,13){\tiny $y$}
      \put(0,30){\line(1,0){90}}
      \put(0,20){\line(1,0){90}}
      \put(90,20){\line(0,1){10}}
      \put(80,20){\line(0,1){10}}
      \put(63,21.5){$\cdots$}
      \put(60,20){\line(0,1){10}}
      \put(50,10){\line(0,1){20}}
      \put(0,10){\line(1,0){50}}
      \put(40,10){\line(0,1){20}}
      \put(16.5,21.5){$\cdots$}
      \put(16.5,11.5){$\cdots$}
      \put(10,10){\line(0,1){20}}
      \put(0,10){\line(0,1){20}}
    }
  \end{picture}
\end{equation}
(where again the length of the first row is greater than or equal to
the length of the second).

\begin{proof}[Proof of Lemma~\bref{lemma:2nd-step}]
  The first three conditions in \eqref{E0-def} are just the definition
  of the space~$\hat\cE$.  Generic elements of $\hat\cE$ are of the
  form
  \begin{gather}
    \phi_0(y,a^{\dagger},\cC)+ b^\dagger\phi_1(y,a^{\dagger},\cC),
  \end{gather}
  where $\phi_0$ and $\phi_1$ satisfy these conditions separately.  We
  write $\hat\cE=\hat\cE_0\oplus\hat\cE_1$ accordingly.
  
  The BRST operator $\hat\brst^{\T}_{-1}$ annihilates $\phi_0$
  trivially, and to identify a unique representative in the
  cohomology, we must fix the freedom of changing $\phi_0$ by a
  coboundary.  The space $\hat\cE_0$ decomposes into a direct sum of
  finite-dimensional representations of the $s\ell(2)$ algebra
  in~\eqref{the-sl2}.  (We recall that the representations are
  finite-dimensional because for any monomial $p$, there exist two
  positive integer numbers $n$ and $m$ such that $(\yS^\dagger)^n p=0$
  and $(\bar\yS^\dagger)^m p=0$.)  In each irreducible summand, there
  exists a unique vector (the lowest-weight vector) that is not in the
  image of $\yS^\dagger$, and it is singled out by the
  condition~$\bar\yS^\dagger\phi_0=0$.  This gives the statement
  about~$H_0$.  The statement about~$H_1$ is obvious.  \end{proof}

The space $\hat\cE$ can be decomposed as
\begin{gather} \label{4.30}
  \hat\cE=\cE_0\oplus\cE_1\oplus \tilde\cF\oplus\tilde\cG,
\end{gather}
where $\tilde\cG=\im\hat\brst^{\T}_{-1}\subset\hat\cE_0$.
Furthermore, exchanging the roles of $\yS^\dagger$ and
$\bar\yS^\dagger$ in the proof of Lemma~\bref{lemma:2nd-step}, we can
also show that $\tilde \cF$ can be taken as $\tilde
\cF=b^\dagger\bar\yS^\dagger\hat\cE\subset \hat\cE_1$, which we do in
what follows.  We also note that because the operators defining the
direct summands in~\eqref{4.30} have definite homogeneity degrees in
both $y^\mu$ and $a^{\dagger\mu}$, the projectors to the summands
preserve the bidegree $(m,n)$ in $(a^{\dagger\mu},y^{\mu})$.

To complete the reduction to the unfolded formalism, in accordance
with Proposition~\bref{prop:parent2unfold}, we next calculate the
reduction of the BRST operator $\hat\brst^\T$ given in
\eqref{brstt-red1} to~$\cE$.

We only need to compute the reduced differential $\tilde\sigma$,
see~\eqref{it-follows}.  By Lemma~\bref{lemma:2nd-step}, the
cohomology of $\brst^{\T}_{-1}$ is concentrated only in degrees $0$
and $1$, and therefore (see the paragraph
following~\eqref{accordingly}), \ $\tilde\sigma$ becomes
\begin{equation}
  \tilde\sigma=\st{\cE\cE}{\sigma}
  - \st{\cE\tilde\cF}{\sigma}\rho\st{\tilde\cG\cE}{\sigma},
\end{equation}
where $\rho\map\tilde\cG\to\tilde\cF$ is the inverse of
$\imath\,\st{\tilde\cG\tilde\cF}{\hat\brst^{\T}_{-1}}
=\yS^\dagger\dl{b^\dagger}$.  Moreover, the degree of
$\st{\cE\tilde\cF}{\sigma}\rho\st{\tilde\cG\cE}{\sigma}$ is $1$, which
implies that $\st{\cE\tilde\cF}{\sigma}\rho\st{\tilde\cG\cE}{\sigma}$
is nonvanishing only as a map from $\cE^{0}$ to~$\cE^{1}$. Along with
the homogeneity degree in $b^\dagger$, we now use the degrees in
$a^{\dagger\mu}$ and $y^\mu$.  Namely, $\hat\brst^{\T}_{-1}$ on
$\hat\cE$ is of bidegree $(1,-1)$ in $(a^{\dagger},y)$ and therefore
its inverse must also be homogeneous, and of bidegree $(-1,1)$.
Because $\sigma$ is of bidegree $(0,-1)$, it follows that
$\st{\cE\tilde\cF}{\sigma}\rho\st{\tilde\cG\cE}{\sigma}$ has bidegree
$(-1,-1)$.  But $\cE_{0}$ contains monomials with the number of $y$'s
greater than or equal to the number of $a^\dagger$'s, while~$\cE_{1}$
contains monomials with the number of $y$'s less than or equal to the
number of~$a^{\dagger}$'s.  We therefore conclude that
$\st{\cE\tilde\cF}{\sigma}\rho\st{\tilde\cG\cE}{\sigma}$ is
nonvanishing only on monomials represented by rectangular Young
tableaux.  These monomials belong to the intersection of the kernels
of $h$, $S^\dagger$, and $\bar S^\dagger$ (and are invariants of the
$s\ell(2)$ algebra in~\eqref{the-sl2}).

For a monomial $\phi \in \cE_{0}$ corresponding to a rectangular Young
tableaux,
\begin{equation} \label{4.32}
  \rho\,\st{\tilde\cG\cE}{\sigma}\phi
  =b^\dagger\bar\yS^\dagger\sigma\phi.
\end{equation}
Indeed, using $\yS^\dagger\bar\yS^\dagger\phi=0$ and
$\commut{\yS^\dagger}{\bar\yS^\dagger}\phi=h\phi=0$, we obtain that
$\bar\yS^\dagger\yS^\dagger\phi=0$, which implies $\yS^\dagger\phi=0$
because no polynomials in $\cE_0$ can be in the image of
$\yS^\dagger$.  Applying then
$\imath\st{\hat\cE\hat\cE}{\brst^{\T}_{-1}}=\yS^\dagger\dl{b^\dagger}$
to both sides and using $\bar\yS^\dagger\phi=\yS^\dagger\phi=h\phi=0$,
we see that Eq.~\eqref{4.32} is equivalent to
$\st{\tilde\cG\cE}{\sigma}\phi=\sigma\phi$. (We here need the explicit
choice of $\tilde\cF$ made above, which guarantees that
$\imath\st{\hat\cE\hat\cE}{\brst^{\T}_{-1}}$ is invertible).  This
last equation is satisfied because for $\phi\in \cE_0$ corresponding
to a rectangular Young tableau, $\sigma\phi=\dl{b^\dagger}\yS^\dagger
\bar\yS^\dagger\sigma\phi\in\tilde\cG$.

Let $\Pi^\cR$ be the projector on the subspace of polynomials in
$\cE_0$ corresponding to rectangular Young tableaux (see
Appendix~\bref{sec:C} for an explicit expression).  Applying $\sigma$
to the right-hand side of~\eqref{4.32} and defining
$\bar\sigma=-\cC^\mu\ffrac{\dd}{\dd a^{\dagger\mu}}$, we then see that
for any $\phi\in\cE_0$,
\begin{equation} \label{eq:srs}
  \st{\cE\tilde\cF}{\sigma}\rho\st{\tilde\cG\cE}{\sigma}\phi
  =b^\dagger\sigma\bar\sigma \,\Pi^\cR\phi,
\end{equation}
where the right-hand side of~\eqref{eq:srs} belongs indeed to $\cE_1$
because it is annihilated by~$\yS^\dagger$.  Finally, the reduced BRST
differential is given by
\begin{equation}
  \label{eq:3}
  \imath\tilde\brst^\T\bundle{\phi}=[\derham+\st{\cE\cE}{\sigma}
  - b^\dagger\sigma\bar\sigma \,\Pi^\cR]\bundle{\phi},\quad
  \bundle{\phi}\in \Gamma(\bundle{\cE}).
\end{equation}
In order to obtain an explicit expression for $\st{\cE\cE}{\sigma}$,
we need the projector $\Pi^{\cE_0}$, whose explicit expression is
given in Appendix~\bref{sec:C}. (The projector $\Pi^{\cE_1}$ is not
needed because $\sigma$ restricts to $\cE_1$.)

We write ${\tilde\Psi}^{\T (0)}$ for the ghost-number-zero part of the
string field $({\hat\Psi}^{\T})^{\cE}$ associated with $\cE$.  Its has
the form
\begin{equation}
  {\tilde\Psi}^{\T (0)}=\tilde\Psi_0 + b^\dagger \tilde\Psi_1,
  \qquad \tilde\Psi_1=\cC^\mu\tilde\Psi_{1\mu},
\end{equation}
where $\tilde\Psi^0$ and $b^\dagger\tilde\Psi_{1\mu}$ are associated
with the respective spaces $\cE_0$ and $\cE_1$.  The corresponding
equations of motion are a reformulation of the unfolded form of the
Fronsdal equations~\cite{Lopatin:1988hz}.\footnote{The authors are
  grateful to M.~Vasiliev for drawing their attention to a problem in
  an earlier attempt to derive equations \eqref{4.35}--\eqref{4.36}.}
\begin{prop}[unfolded Fronsdal equations]
  \label{prop:F-unfolded}
  The equations of motion $\tilde\brst^{\T} \tilde\Psi^{\T (0)}=0$ are
  given by
  \begin{align}
    (\derham + \Pi^{\cE_0}\sigma)\tilde\Psi_0&=0,\label{4.35}\\
    (\derham + \sigma)\tilde\Psi_1&=-\sigma\bar\sigma
    \,\Pi^\cR\tilde\Psi_0.
    \label{4.36}
  \end{align}
\end{prop}

Under the gauge transformations $\delta\tilde\Psi^{\T
  (0)}=\tilde\brst^\T\tilde\Psi^{\T (1)}$ with $\tilde\Psi^{\T
  (1)}=-\imath b^\dagger\tilde\Lambda^{}$, the component fields in
$\tilde\Psi_0$ are invariant and can therefore be interpreted as
generalized curvatures, while the component fields in $\tilde\Psi_1$,
to be interpreted as generalized gauge fields, transform~as
\begin{equation}
  \delta \tilde\Psi_1
  =(\derham + \sigma)\tilde\Lambda^{},
\end{equation}
where the component fields in $\tilde\Lambda^{}$, which are of ghost
number $1$, are to be replaced with gauge parameters.  Because the
component fields $\tilde\Psi^{\T (1)}$ are associated with states at
ghost number $-1$, the tensor structure of the component fields of
$\tilde\Lambda^{}$ and hence also of the gauge parameters is described
by~\eqref{eq:YT2}.

\section{Discussion}
\subsubsection*{Summary}
The main result in this paper can be summarized by the diagram
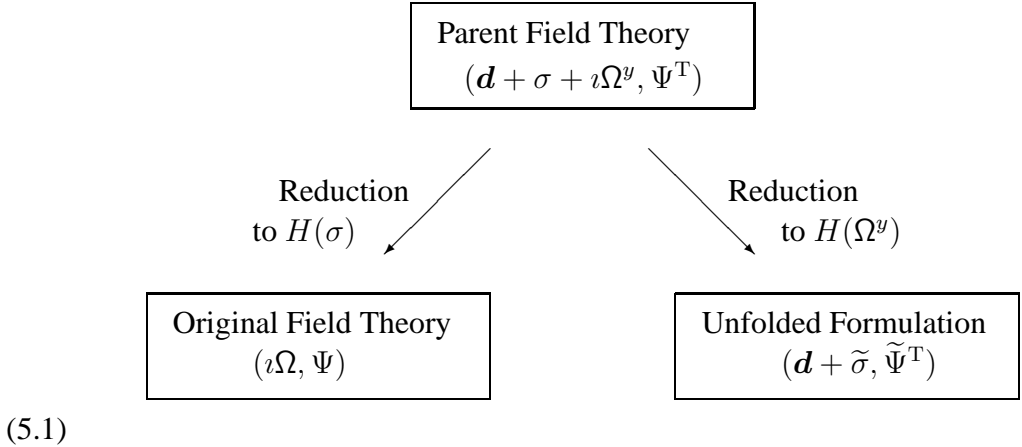
\begin{equation}
  \begin{picture}(200,170)
    \put(-30,0){\put(80,150){Parent Field Theory}
      \put(90,133){$(\derham+\sigma+\imath\brst^y,\Psi^{\T})$}
      \put(70,165){\line(1,0){130}}
      \put(70,125){\line(1,0){130}}
      \put(70,125){\line(0,1){40}}
      \put(200,125){\line(0,1){40}}
      \put(-20,40){Original Field Theory}
      \put(10,25){$( \imath\brst,\Psi)$}
      \put(-100,-110){
        \put(70,165){\line(1,0){130}}
        \put(70,125){\line(1,0){130}}
        \put(70,125){\line(0,1){40}}
        \put(200,125){\line(0,1){40}}}
      \put(180,40){Unfolded Formulation}
      \put(210,25){$(\derham+\tilde\sigma,\tilde\Psi^{\T})$}
      \put(100,-110){
        \put(70,165){\line(1,0){130}}
        \put(70,125){\line(1,0){130}}
        \put(70,125){\line(0,1){40}}
        \put(200,125){\line(0,1){40}}}
      \put(160,110){\vector(1,-1){40}}
      \put(100,110){\vector(-1,-1){40}}
      \put(190,90){Reduction}
      \put(210,75){to $H(\brst^y)$}
      \put(20,90){Reduction}
      \put(10,75){to $H(\sigma)$}}
  \end{picture}
\end{equation}

\subsubsection*{Curved backgrounds}
At the first-quantized level, the construction of the extended
first-order system is a simplified and adapted version of Fedosov
quantization, which has been developed in order to study quantization
of curved manifolds.  This is the reason why the construction of the
extended system can naturally be done also for a quantum system on a
curved background.  Furthermore, using the full power of the Fedosov
method provides a framework to study the extension of quantum systems
from flat to curved backgrounds.  We plan to discuss these matters in
more details elsewhere.

\subsubsection*{Symmetries}
We also comment on how global symmetries and their representations
enter the parent system.  The standard setting
\cite{Vasiliev:1999ba,Shaynkman:2004vu} of the unfolded formalism
consists in specifying a Lie algebra~$\algg$ of global symmetries and
its representation.  The fields take values in this representation and
the equations have the form of the covariant constancy condition with
respect to a flat $\algg$-connection.  These data inherently occur in
the parent system: the representation is given by
$H^0(\brst^y,\cH^{\T})$ and the enveloping algebra of~$\algg$ in this
representation is given by $H^0(\commut{\brst^y}{\cdot\,})$.

This realization of the unfolded setting in the BRST theory could be
expected for the following reasons.  At the first-quantized level,
elements of $H^0(\commut{\brst}{\cdot\,})$ are naturally interpreted
as \textit{observables} of the corresponding system.  At the level of
the associated field theory, these elements then determine linear
global \textit{symmetries}.

In the example of the Klein--Gordon equations (see
Sec.~\bref{sec:red-unfolded}), the BRST cohomology
$H^0(\brst^y,\cH^{\T})$ can be calculated and agrees with the algebra
obtained in~\cite{Eastwood:2002su}.

\enlargethispage{6pt}

\subsubsection*{Interactions}
In the Lagrangian case, the appropriate deformation theory for
studying consistent interactions has been developed in
\cite{Barnich:1993vg} (see also \cite{Henneaux:1997bm}).  It is based
on the graded differential algebra composed of local functionals in
the fields and antifields graded by ghost number, the field--antifield
antibracket, and the Batalin--Vilkovisky master action.  In the
non-Lagrangian context, according to Sec.~\bref{sec3.1}, the field
theory is described by a BRST differential $s$ that is a nilpotent
evolutionary vector field on the space of fields and their
derivatives.  The graded differential algebra on which deformation
theory is based is then composed of the space of evolutionary vector
fields in the fields and their derivatives, graded again by ghost
number, the commutator bracket $[\cdot,\cdot]$ for evolutionary vector
fields and the vector field $s$.  As usual~\cite{Gerstenhaber:1964},
it follows from
\begin{equation}
  \half[s+g\st{(1)}{s}+\dots,s+g\st{(1)}{s}+\dots]=0,
\end{equation}
where $g$ is some deformation parameter, that first-order nontrivial
interactions $\smash{\st{(1)}{s}}$ are representatives of the
cohomology of the adjoint action of $s$ in ghost number $1$, while
obstructions to first-order deformations are controlled by the bracket
induced in the cohomology.  Because the adjoint cohomology of $s$ in
the space of evolutionary vector fields is invariant under the
elimination or introduction of generalized auxiliary fields, so are
the nontrivial consistent interactions.

If the expansion is in terms of homogeneity in the fields, for
instance
\begin{equation}
  \st{(1)}{s}=
  \st{(1)}{K^A}_{BC}(\psi^B,\psi^C)\dl{\psi^A}+
  \partial_\mu(\st{(1)}{K^A}_{BC}(\psi^B,\psi^C)\dl{\psi^A_{,\mu}})
  +\dots,
\end{equation}
with $\smash{\st{(1)}{K^A}_{BC}(\psi^B,\psi^C)}$ denoting local
functions of homogeneity~$2$ in the fields and their derivatives, the
deformation problem can be entirely reformulated at the
first-quantized level: the existence of a consistent deformation
corresponds to the existence of multilinear graded symmetric
differential operators $K:\Gamma(\bundle{\cH})^{\otimes n}\rightarrow
\Gamma(\bundle{\cH})$ of ghost number $1$ for $n\geq 2$ that combine
with $\brst$ into an $L_\infty$ algebra, see,
e.g.,~\cite{Lada:1993wc}.

Of course, after having found a consistent deformation of the
differential $s$, additional constraints need to be satisfied in order
for this deformation to be associated with a (real) Lagrangian gauge
field theory.

\subsubsection*{Acknowledgments}
We are grateful to K.~Alkalaev, G.~Bonelli, N.~Boulanger, B.L.~Feigin,
E.B.~Feigin, O.~Sheynkman, P.~Sundell, and especially M.A.~Vasiliev
for useful discussions.  MG and IYuT thank M.~Henneaux for kind
hospitality at Universit\'e Libre de Bruxelles.  The work of GB and MG
is supported in part by a ``P{\^o}le d'Attraction Interuniversitaire''
(Belgium), by IISN-Belgium, convention 4.4505.86, and by the European
Commission RTN program HPRN-CT00131, in which the authors are
associated to K.~U.~Leuven.  GB is also supported by Proyectos
FONDECYT 1970151 and 7960001 (Chile); MG and AMS are supported in part
by the INTAS (Grant 00-00262), the RFBR Grant 04-01-00303,
02-01-00930, and by the Grant LSS-1578.2003.2; IYuT is supported in
part by the RFBR Grant 02-02-16944 and by the Grant LSS-1578.2003.2.


\appendix
\section{Structure of some polynomial $sp(4)$ representations}
\label{sec:A}
For the operators $\Box$, $\yS$, and $T$ involved in the BRST operator
$\brst^{\T}_{-1}$ in~\eqref{T-split}, we here show a property that
allows calculating the cohomology of~$\brst^{\T}_{-1}$ acting on the
space $\mathscr{P}_{\dmn}(y,a^\dagger)$ of polynomials in $y^\mu$ and
$a^{\dagger\mu}$, $\mu=1,\dots,\dmn$ (tensored with the appropriate
ghosts).  This can be done by standard representation-theory methods.

The space $\mathscr{P}_{\dmn}(y,a^\dagger)$
carries representations of two Lie algebras, the Lorentz algebra
acting in the standard way and the $sp(4)$ algebra whose Chevalley
basis generators are represented as
\begin{equation}\label{all-sp4}
  \begin{alignedat}{4}
    T&= \eta^{\mu\nu}\dl{a^{\dagger \mu}} \dl{a^{\dagger \nu}}, &\quad
    \yS^\dagger& = a^{\dagger \mu}\dl{y^\mu}, &\quad
    \yS&=\eta^{\mu\nu}\dl{a^{\dagger \mu}} \dl{y^\nu}, &\quad
    \Box&= \eta^{\mu\nu} \dl{y^\mu} \dl{y^\nu},\\
    h'&=-a^{\dagger\nu} \dl{a^{\dagger \nu}}-\ffrac{\dmn}{2}, &
    h&=a^{\dagger\mu} \dl{a^{\dagger \mu}} - y^\mu
    \dl{y^\mu},\kern-60pt
    \\
    \bar T&=-\ffrac{1}{4} \eta^{\mu\nu}a^{\dagger}_\mu
    a^{\dagger}_\nu, \quad& \bar\yS^\dagger&=y^\mu \dl{a^{\dagger
        \mu}}, & \bar\yS&=y^\mu a^\dagger_\mu, \qquad& \bar\Box&=
    y^\mu y^\nu\eta_{\mu \nu}.
  \end{alignedat}
\end{equation}
These two algebras commute, and by Howe duality~\cite{Howe, Howe1},
$\mathscr{P}_{\dmn}(y,a^\dagger)$ decomposes into a direct sum of
irreducible $sp(4)$ representations.  We can therefore restrict the
analysis to an individual irreducible $sp(4)$ representation.

With the space $\mathfrak{h}^*$ dual to the Caratan subalgebra of
$sp(4)$ identified with the standard plane $\fR^2$, the $sp(4)$ root
diagram can be represented as
\begin{equation}
  \begin{gathered}
    \xymatrix@=40pt{
      *{}&*{}&*{}\\
      *{}&*{}%
      \ar[0,-1]_(.8){\bar\yS^\dagger} \ar[-1,-1]_(.7){T}
      \ar[-1,0]_(.9){\yS} \ar[-1,1]_(.9){\Box}
      \ar[0,1]^(.9){\yS^\dagger} \ar[1,1]^(.9){\bar T}
      \ar[1,0]^(.8){\bar\yS} \ar[1,-1]_(.8){\bar\Box\!\!}
      &*{}\\
      *{}&*{}&*{}}
  \end{gathered}
\end{equation}
There is the ``horizontal'' $s\ell(2)$ algebra generated by
$e=\yS^\dagger$, $f=\bar\yS^\dagger$, and $h$, with the commutation
relations given by
\begin{equation}\label{the-sl2}
  \commut{e}{f}=h,
  \quad
  \commut{h}{e}=2e,
  \quad
  \commut{h}{f}=-2f.
\end{equation}

Each irreducible $sp(4)$ representation $\mathscr{I}_\lambda$
occurring in the decomposition of $\mathscr{P}_{\dmn}(y,a^\dagger)$ is
a highest-weight representation, with the highest-weight vector
defined by the annihilation conditions $\yS^\dagger v_\lambda=T
v_\lambda=0$ (which clearly imply $\Box v_\lambda=0=Tv_\lambda$).  The
condition $S^\dagger v_\lambda=0$ also implies that the corresponding
polynomial in $(a^{\dagger\mu},y^\mu)$ has the symmetry of the Young
tableaux
\smash{\unitlength=.8pt%
  \begin{picture}(110,25)
    \put(15,-16){
      \put(-12,23){\tiny $a^\dagger$}
      \put(-12,13){\tiny $y$}
      \put(0,30){\line(1,0){90}}
      \put(0,20){\line(1,0){90}}
      \put(90,20){\line(0,1){10}}
      \put(80,20){\line(0,1){10}}
      \put(63,21.5){$\cdots$}
      \put(60,20){\line(0,1){10}}
      \put(50,10){\line(0,1){20}}
      \put(0,10){\line(1,0){50}}
      \put(40,10){\line(0,1){20}}
      \put(16.5,21.5){$\cdots$}
      \put(16.5,11.5){$\cdots$}
      \put(10,10){\line(0,1){20}}
      \put(0,10){\line(0,1){20}}
    }
  \end{picture}}.  In particular,  if the polynomial is of bidegree
$(m,n)$ in $(a^{\dagger\mu},y^\mu)$, it follows that $m\geq n\geq0$.
We also let $\mathscr{K}_\lambda$ denote the representation of the
horizontal $s\ell(2)$ subalgebra generated from~$v_\lambda$.  It is
finite-dimensional because any monomial $p$ satisfies the conditions
$(\yS^\dagger)^n p=0$ and $(\bar\yS^\dagger)^m p=0$ for two positive
integers~$n$ and~$m$.

Each $\mathscr{I}_\lambda$ is in fact a quotient of (or coincides
with) the \textit{generalized Verma module~$\mathscr{W}_\lambda$}
induced from a finite-dimensional representation
$\mathscr{K}_\lambda$.  A generalized Verma module can be simply
defined in our setting as the space of polynomials in $\bar\Box$,
$\bar\yS$, and $\bar T$ with coefficients
in~$\mathscr{K}_\lambda$.\footnote{More formally, let $\algp$ be the
  parabolic subalgebra in $sp(4)$ generated by $\bar\yS^\dagger$, $T$,
  $\yS^\dagger$, and $h'$ and let $\mathscr{K}_{\lambda}$ be a
  representation of~$\algp$ where $T$, $\yS$, and $\Box$ act
  trivially.  Then $\mathscr{W}_\lambda=\mathscr{U}(sp(4))
  \tensor_{\mathscr{U}(\algp)}\mathscr{K}_{\lambda}$.}  Therefore,
instead of the highest-weight vector in an ordinary Verma module,
there is a representation of the horizontal $s\ell(2)$ algebra in a
generalized Verma module.

We next follow the standard analysis~\cite{Dixmier} that
consists in studying the shifted Weyl group action and finding the
weights in the Weyl group orbit that are in the fundamental Weyl
chamber.  To describe the representation weights, we introduce
elements~$\varepsilon_1$ and~$\varepsilon_2$ of the orthonormal basis
in~$\mathfrak{h}^*$. In the above root diagram, $\varepsilon_1$
and~$\varepsilon_2$ are identified as the roots corresponding to~$\yS$
and~$\yS^\dagger$ respectively.  The two positive simple roots
$\alpha$ and $\alpha'$ are given by
\begin{gather}
  \alpha=\varepsilon_2,\qquad \alpha'=\varepsilon_1-\varepsilon_2\,,
\end{gather}
and hence are the roots corresponding to $\yS^\dagger$ and $T$
respectively.
Half the sum of positive roots is given by
$\rho=\frac{3}{2}\varepsilon_1+\frac{1}{2}\varepsilon_2$.
The weights are parameterized as
\begin{equation}\label{the-weights}
  \begin{gathered}
    \lambda=a\varepsilon_1 + b\varepsilon_2,\\
    a=-\half(m+n+\dmn),\qquad b=\half(m-n),
  \end{gathered}
\end{equation}
where we recall that $m\geq n\geq 0$.

The Weyl group action shifted by $-\rho$ is generated by the two
mappings
\begin{equation}
  (a,b)\mapsto(a,-b-1)\quad\text{and}\quad(a,b)\mapsto(b-1,a+1).
\end{equation}
The fundamental Weyl chamber $\wch_{-\rho}$ (translated by $-\rho$) is
determined by the conditions
\begin{equation}
  a\varepsilon_1 + b\varepsilon_2\in\wch_{-\rho}
  \Longleftrightarrow
  (b+\half\geq0,\quad a-b+1\geq0).
\end{equation}
These conditions are to be examined for each element of the Weyl group
orbit of~\eqref{the-weights}, with $m\geq n\geq0$.  With the dominant
weight found, the Bruhat order on the Weyl group induces an order on
the weights.  For \textit{generalized} Verma modules, the analysis
differs from the standard in that only those weights are kept in the
orbit whose component ``along the horizontal subalgebra'' (the
$\varepsilon_2$-component in our case) is dominant (see,
e.g.,~\cite{mazorchuk-generalized} for details in the case of
generalized Verma modules).

The Bruhat order induced on the representation weights depends on the
dimension $\dmn$, and analysis shows that for $\dmn=1,2$, some of the
representations in the decomposition are quotients of the
corresponding generalized Verma modules over nontrivial submodules,
but for $\dmn\geq3$, they necessarily coincide with the generalized
Verma modules.

For $\dmn\geq3$, this has the following implication for the properties
of the $\Box$, $\yS$, and $T$ operators.  As in the proof of
Lemma~\bref{lemma:1st-step}, let $\spaceN$ be the subspace of
polynomials annihilated by $T$ and~$\Box$ in a given irreducible
representation $\mathscr{I}_\lambda$ and $\yS_\spaceN$ be the
restriction of~$\yS$ to~$\spaceN$.  Then $\im \yS_\spaceN=\spaceN$.

Indeed, in an irreducible representation~$\mathscr{I}_\lambda$, let
$\mathscr{I}^\ell_\lambda$ denote the subspace spanned by elements of
the form $\bar\Box^{\ell_1}\bar{T}^{\ell_2}\bar{S}^{\ell_3}\chi$,
where $\chi\in \mathscr{K}_{\lambda}$ and $\ell_1+\ell_2+\ell_3=\ell$
with $\ell,\ell_i\geq0$.  The property $\im \yS_\spaceN=\spaceN$ is
equivalently reformulated as the property that all subspaces
$\mathscr{N}^\ell_\lambda=\ker T\cap \ker \Box \cap
\mathscr{I}^\ell_\lambda$ are isomorphic for $\ell\geq 0$.  On the
other hand, we can expand $\phi^\ell\in \mathscr{I}^\ell_{\lambda}$
into the $T$- and $\Box$-traceless part and the respective traces~as
\begin{equation}
  \phi^\ell=\phi^\ell_{0,0}
  +\sum_{\ell^{\,\prime}=1}^{\ell}\,
  \mathop{{\sum_{\ell_1,\ell_2\geq0}}}_{\ell_1 + \ell_2=\ell}
  \bar\Box^{\ell_1} \bar T^{\ell_1} \phi^\ell_{\ell_1,\ell_2},
\end{equation}
where $\phi^\ell_{\ell_1,\ell_2}\in\spaceN^{\ell-\ell_1-\ell_2}$. \ 
This immediately gives the formula
\begin{equation}\label{max-formula}
  \dim \mathscr{I}^\ell_\lambda
  = \sum_{\ell^{\,\prime}<\ell}(\ell'+1)
  \dim\mathscr{N}^{\ell-\ell^{\,\prime}}_\lambda.
\end{equation}

But as we have shown, $\mathscr{I}_\lambda=\mathscr{W}_\lambda$.  An
elementary calculation of the character of the generalized Verma
module then shows that
\begin{equation}\label{eq:dim}
  \dim \mathscr{I}^\ell_\lambda=
  \dim \mathscr{W}^\ell_\lambda = (\dim\mathscr{K}_{\lambda})\,
  \ffrac{(\ell+1)(\ell+2)}{2}.
\end{equation}
Comparing~\eqref{max-formula} and~\eqref{eq:dim}, we conclude that
$\dim\mathscr{N}^\ell_\lambda=\dim\mathscr{K}_\lambda$, independently
of~$\ell$, and therefore the $\mathscr{N}^\ell_\lambda$ spaces are
isomorphic for all~$\ell$.  As noted above, this is equivalent to the
desired property of~$\yS_\spaceN$.

\subsubsection*{Remarks}\mbox{}

(i)~The BRST operator $\brst^{\T}_{-1}$ in~\eqref{T-split} evaluates
the cohomology of the Abelian subalgebra in $sp(4)$ generated by $T$,
$\yS$, and $\Box$.  The occurrence of generalized Verma modules may be
traced to the fact that the annihilation conditions with respect to
these three operators constitute the highest-weight conditions in the
$sp(4)$ generalized Verma modules.  The ``opposite'' operators $\bar
T$, $\bar\yS$, and $\bar\Box$ are then the creation operators in the
generalized Verma modules.

(ii)~In the language of the standard $sp(4)$ Verma modules, the
statement about the structure of $\mathscr{I}_\lambda$ means that for
$\dmn\geq 3$, each irreducible representation $\mathscr{I}_\lambda$ is
a quotient of the Verma module $\mathscr{V}_\lambda$ with respect to
only one singular vector, given by
$(\bar\yS^\dagger)^{2b+1}v_\lambda$, but for $\dmn=1,2$, some of the
representations $\mathscr{I}_\lambda$ are given by a quotient of
$\mathscr{V}_\lambda$ over \textit{two} singular vectors.

\section{Projectors}
\label{sec:C}
The horizontal $s\ell(2)$ algebra in~\eqref{the-sl2}, which played a
crucial role in the proof of Lemma~\bref{lemma:2nd-step}, is also
helpful in writing the projectors~$\Pi^\cR$ and~$\Pi^{\cE_0}$ that
occur in~\eqref{eq:srs}--\eqref{4.36}.  We first consider the
projector
\begin{equation}
  \Pi^{\cE_0}:\hat\cE_0\to\cE_0,
\end{equation}
where $\hat\cE_0$ is a direct sum of finite-dimensional irreducible
$s\ell(2)$ representations and $\cE_0$, defined in
Lemma~\bref{lemma:2nd-step}, is the subspace of lowest-weight vectors
(those annihilated by the~$f$ generator of~$s\ell(2)$).  We
have~\cite{Zhelob}
\begin{equation}\label{B2}
  \Pi^{\cE_0}=\sum_{n\geq0}
  \ffrac{1}{n!}\frac{1}{\prod\limits_{j=1}^n(h-j-1)}\,
  e^nf^n
  =\prod_{n\geq1}\Bigl(1+\ffrac{ef}{n(h-n-1)}\Bigr).
\end{equation}
It is easy to verify that $f\Pi^{\cE_0}=\Pi^{\cE_0} e=0$: for example,
for $f\Pi^{\cE_0}$, the relevant commutation relations are
\begin{equation}
  [f,e^n]=-n(h-n+1)e^{n-1}
\end{equation}
and
\begin{equation}
  f F(h)= F(h+2)f
\end{equation}
for ``any'' function~$F$.  Therefore, taking the $n$th term
in~\eqref{B2}, we have
\begin{multline}
  f\cdot\ffrac{1}{n!}\,\frac{1}{\prod\limits_{j=1}^n(h-j-1)}\,e^nf^n\\*
  =\ffrac{1}{n!}\,\frac{1}{\prod\limits_{j=1}^n(h-j+1)}\,
  (-n(h-n+1)e^{n-1}f^n + e^nf^{n+1})\\*
  =-\ffrac{1}{(n-1)!}\,\frac{1}{\prod\limits_{j=1}^{n-1}(h-j+1)}\,e^{n-1}
  f^n +
  \ffrac{1}{n!}\,\frac{1}{\prod\limits_{j=1}^n(h-j+1)}\,e^nf^{n+1},
\end{multline}
which gives zero after summation over~$n$.

Next, the operator~$\Pi^\cR$ occurred in Lemma~\bref{lemma:2nd-step}
as the projector on rectangular two-row Young tableaux.  In terms of
the $s\ell(2)$ algebra, it is the projector on $s\ell(2)$ invariants
(the trivial representation) and can therefore be written as
\begin{equation}
  \Pi^\cR
  =\prod_{n\geq1}\Bigl(1-\ffrac{C}{n(n+2)}\Bigr)
  =-\ffrac{2\sin(\pi\sqrt{1+C})}{\pi C\sqrt{1+C}},
\end{equation}
where $C=4ef+h^2-2h$ is the $s\ell(2)$ Casimir operator.  This
projector simplifies on the image of~$\Pi^{\cE_0}$:
\begin{equation}
  \Pi^\cR\,\Pi^{\cE_0}=
  \ffrac{2\sin(\pi h)}{\pi h(h+1)(h+2)}\,\Pi^{\cE_0}.
\end{equation}


\providecommand{\href}[2]{#2}\begingroup\raggedright\endgroup
\end{document}